\documentclass[
superscriptaddress,
 amsmath,amssymb,
 aps,
 prl,
 twocolumn,
]{revtex4}
\usepackage{amsmath}
\usepackage{amssymb}
\usepackage{graphicx}
\usepackage{dcolumn}
\usepackage{bm}
\usepackage{xcolor}
\usepackage[export]{adjustbox}
\usepackage{stfloats}

\begin{document}

\preprint{APS/123-QED}

\title{Surface Passivation Method for Super-repellence of Aqueous Macromolecular Condensates}

\author{Andrea Testa}
 \thanks{These authors contributed equally.}
  \affiliation{Department of Materials, ETH Z\"{u}rich, 8093 Z\"{u}rich, Switzerland.}

\author{Hendrik T. Spanke}
 \thanks{These authors contributed equally.}
  \affiliation{Department of Materials, ETH Z\"{u}rich, 8093 Z\"{u}rich, Switzerland.}
  
\author{Etienne Jambon-Puillet}
  \affiliation{Department of Materials, ETH Z\"{u}rich, 8093 Z\"{u}rich, Switzerland.}
  \affiliation{LadHyX, CNRS, Ecole Polytechnique, Institut Polytechnique de Paris, Palaiseau 91120, France}
  
\author{Mohammad Yasir}
  \affiliation{Department of Materials, ETH Z\"{u}rich, 8093 Z\"{u}rich, Switzerland.}
  
\author{Andreas M. K\"{u}ffner}
  \affiliation{Department of Chemistry and Applied Biosciences, Institute for Chemical and Bioengineering, ETH Z\"{u}rich, 8093 Z\"{u}rich, Switzerland.}

\author{Paolo Arosio}
  \affiliation{Department of Chemistry and Applied Biosciences, Institute for Chemical and Bioengineering, ETH Z\"{u}rich, 8093 Z\"{u}rich, Switzerland.}

\author{Eric R. Dufresne}
  \affiliation{Department of Materials, ETH Z\"{u}rich, 8093 Z\"{u}rich, Switzerland.}

\author{Robert W. Style}
  \affiliation{Department of Materials, ETH Z\"{u}rich, 8093 Z\"{u}rich, Switzerland.}
  
\author{Aleksander A. Rebane}
 \email{a.rebane@nyu.edu}
  \affiliation{Department of Materials, ETH Z\"{u}rich, 8093 Z\"{u}rich, Switzerland.}
    \affiliation{Life Molecules and Materials Lab, New York University Abu Dhabi, P.O. Box 129188, Abu Dhabi, United Arab Emirates.}

\date{\today}
             
\begin{abstract}
Solutions of macromolecules can undergo liquid-liquid phase separation to form droplets with ultra-low surface tension.
Droplets with such low surface tension wet and spread over common surfaces such as test tubes and microscope slides, complicating \textit{in vitro} experiments.
Development of an universal super-repellent surface for macromolecular droplets has remained elusive because their ultra-low surface tension requires low surface energies.
Furthermore, nonwetting of droplets containing proteins poses additional challenges because the surface must remain inert to a wide range of chemistries presented by the various amino-acid side-chains at the droplet surface.
Here, we present a method to coat microscope slides with a thin transparent hydrogel that exhibits complete dewetting (contact angles $\theta\approx180^\circ)$ and minimal pinning of phase-separated droplets in aqueous solution.
The hydrogel is based on a swollen matrix of chemically crosslinked polyethylene glycol diacrylate of molecular weight 12 kDa (PEGDA), and can be prepared with basic chemistry lab equipment.
The PEGDA hydrogel is a powerful tool for \textit{in vitro} studies of weak interactions, dynamics, and internal organization of phase-separated droplets in aqueous solutions.
\end{abstract}

\maketitle

\section{Introduction}
Aqueous solutions of macromolecules can undergo phase separation, forming droplets enriched in one or more components.
This phenomena has garnered widespread interest in recent years thanks to the discovery that membraneless organelles inside cells appear to be phase separated droplets of proteins and/or nucleic acids (bio-condensates) \cite{banani2017biomolecular, boeynaems2018protein}. 
Examples include P granules \cite{brangwynne2009germline}, P bodies \cite{Kato2012CellfreeFO, Nott2015PhaseTO}, stress granules \cite{molliex2015phase}, and the nucleolus \cite{Brangwynne2011ActiveLB}.
The material properties and internal organization of bio-condensates are thought to be important for biological function \cite{SchusterLLPSReview2021,LaFontaineNucleolus2020}, and aberrations therein have been implicated in neurodegenerative diseases \cite{ShinLLPS2017, ALSFUS2015}.

\begin{figure*}
    \centering
   \includegraphics[width=\textwidth]{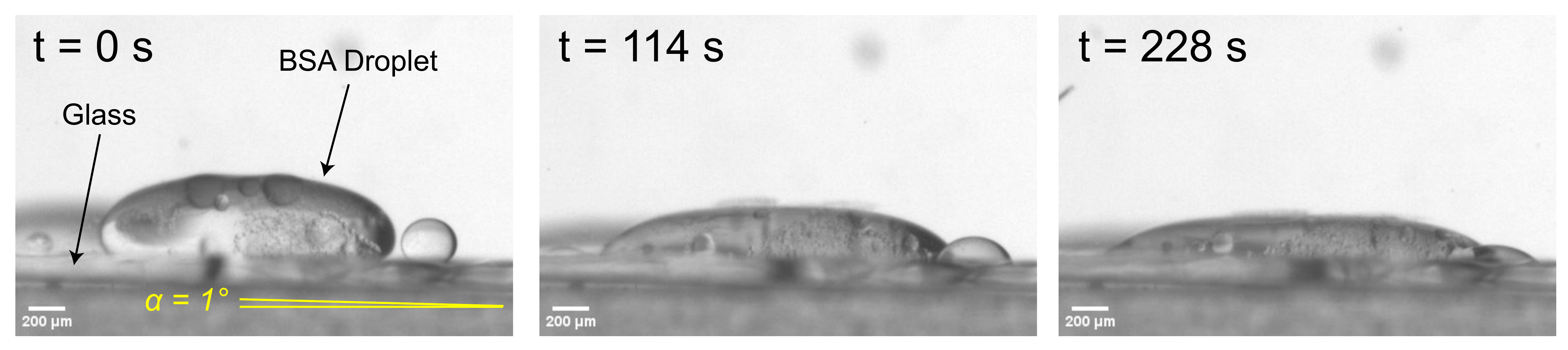}
    \caption{\textit{Supporting Figure: BSA droplets wetting and spreading on bare glass.} 
    }
    \label{supp_fig:bsa_sliding_on_glass}
\end{figure*}

A notable feature of bio-condensates is their extremely low surface tension, with reported values as low as $\gamma\sim1~\mu\mathrm{N}/\mathrm{m}$ \textit{in vivo} \cite{brangwynne2009germline,BrachaProbing2019,feric2016coexisting} and \textit{in vitro} \cite{feric2016coexisting,WANGaspiration20211,ijavi2021surface,FusionOT2021, JawerthOT}.
Such low surface tension results in a strong tendency of bio-condensates to adhere to most common substrates, including walls of test tubes and microscope slides (Fig. S\ref{supp_fig:bsa_sliding_on_glass}) \cite{WettingSurfaceTension1960}, often to the detriment of \textit{in vitro} experiments.
For instance, droplet pinning causes underestimates of surface tension determined from the kinetics of fusion between two sessile droplets on a microscope slide \cite{eggersDrops1999,feric2016coexisting, FusionOT2021}. 
Similarly, wetting may introduce systematic errors to viscosity and viscoelasticity values obtained from passive and active microrheology \cite{ijavi2021surface, JawerthOT}. 
Further challenges arise when including macromolecular crowding in the buffer to more closely mimic the physicochemical properties of the cytoplasm \cite{ellis2001macromolecular}. 
Crowding promotes non-specific interactions that lead to even stronger adhesion of the bio-condensates to microscope slides \cite{hoppe2019non}.
Thus, accurate characterization of bio-condensates \textit{in vitro} requires surface passivation techniques that minimize interactions between the droplets and the experimental substrates.

Numerous surface passivation strategies have been developed to counteract  droplet wetting \cite{Repellent2022}. Microscope slides are typically coated with bovine serum albumin (BSA) \cite{chen2020liquid, lin2015formation}, fluorinated fluids \cite{feric2016coexisting, jeon2018salt}, slippery omniphobic covalently attached liquids (SOCAL) \cite{SOCAL2016}, PEG based polymer brushes \cite{Sofia1998PolyethyleneOG, Upadhyayula2012CoatingsOP, kirkness2018modified}, or treated to form a slippery liquid-infused porous surface (SLIPS) \cite{SLIPSnature2011}.
These treatments generally achieve selective dewetting conditions of either aqueous droplets (hydrophobic), non-polar droplets (oleophobic), or certain combinations thereof (omniphobic).
However, the creation of a surface coating that is super-repellent to bio-condensates has remained elusive because they simultaneously present divergent chemistries on their surfaces, ranging from charged or polar to aliphatic or aromatic.
Here, we put forward an optimized hydrogel surface that alleviates this issue by exhibiting complete dewetting of a broad range of phase-separated droplets in aqueous solution, thus allowing detailed quantitative characterizations of unperturbed droplets \textit{in vitro}.

\section{Results and Discussion}
%
\subsection{Experimental design}
We set out to formulate a substrate exhibiting complete dewetting of macromolecular  droplets in aqueous solution (\textit{e.g.} protein droplets).
We focused on optically transparent coatings that could be applied to light microscopy glass slides for imaging, and be synthesized by non-experts using basic chemistry lab equipment. 
Specifically, we chose to investigate the commonly used substrates of bare glass, covalently attached polyethylene glycol brush on glass (PEG-silane), a slippery liquid-infused porous surface (SLIPS) of silicone oil, and a novel coating made of chemically cross-linked polyethylene glycol diacrylate of molecular weight $12~\textrm{kDa}$ (PEGDA) hydrogel.

Previously, we have successfully suppressed depletion-induced adhesion between a microscope glass and giant unilammellar vesicles by coating the glass with a low molecular weight PEGDA ($700~\textrm{Da}$) hydrogel \cite{spanke2020wrapping}.
However, we found that this hydrogel still exhibited partial wetting of protein droplets, likely due to non-specific attractions between PEGDA and protein \cite{testa2021sustained}.
We  postulated that by increasing the molecular weight of the PEGDA, we could achieve a higher volume fraction of water and lower volume fraction of polymer within the hydrogel, thereby making the hydrogel surface more similar to the solvent and reducing adhesive protein-hydrogel interactions.
We therefore reformulated the hydrogel coating with PEGDA $12~\textrm{kDa}$.

Detailed protocols for preparing these surfaces and synthesizing the PEGDA are given in the Experimental Section.
Briefly, we prepared PEG-silane glass slides by submerging glass coverslips in a toluene solution of 3-[methoxy(polyethylene-oxy)propyl]trimethoxysilane (PEG-silane) overnight \cite{ijavi2021surface}. For SLIPS, we spray-coated glass coverslips with hydrophobic particles and then infused the resultant matrix with vinyl-terminated polydimethylsiloxane (silicone oil) by spin-coating \cite{Bradley2019PRL}. 
We prepared the PEGDA hydrogel by pre-treating coverslips with the silane coupling agent 3-(trimethoxysilyl) propylmethacrylate \cite{spanke2020wrapping}. 
We then added a solution of PEGDA 12 kDa and photoinitiator (2-Hydroxy-4’-(2-hydroxyyethoxy)-2-methylpropiophenone) to the pre-treated glass slide, covered it with another glass slide coated with an anti-sticking film (RainX\textsuperscript{\textregistered}), and subsequently cured the hydrogel under UV light. 
The glass sandwich was then opened and immediately used for experiment.
A schematic illustrating the key steps in PEGDA hydrogel preparation is shown in Figure S\ref{fig:pegda_substrate_making}.
\subsection{Static contact angle}
\begin{figure*}[t]
    \centering
    \includegraphics[width=\textwidth]{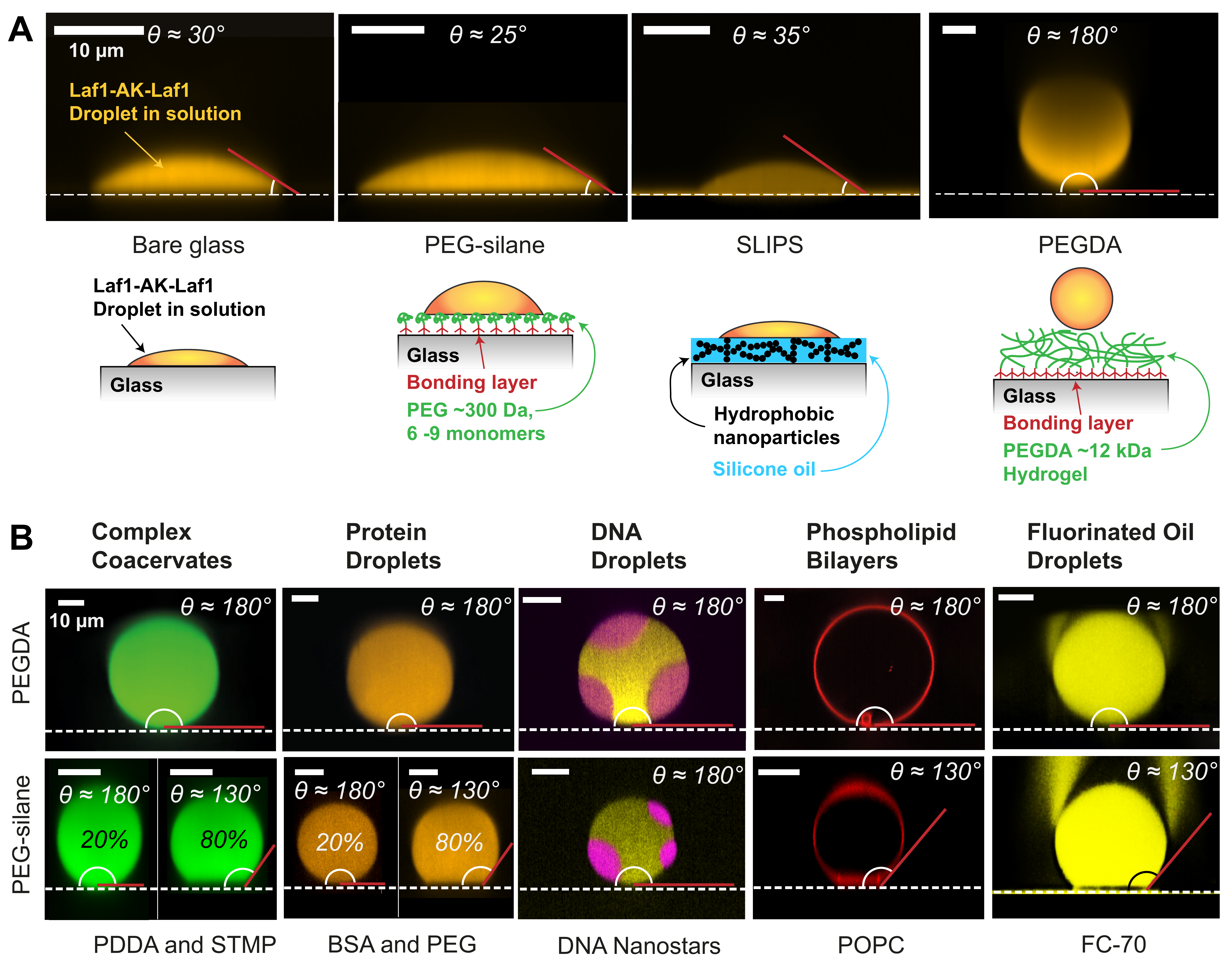}
    \caption{
    \textit{PEGDA hydrogel coating exhibits complete dewetting of droplets in aqueous solution.}
    \textbf{(A)} Top: Side view and static contact angle of fluorescently labelled Laf1-AK-Laf1 droplets (\textit{orange}) resting on different substrates (\textit{white dashed line}). 
    The red line marks the tangent to the droplet surface at the contact line. 
    The white arc is the contact angle, $\theta$, which exhibits partial wetting on glass ($\theta \approx 30^{\circ}$), PEG-silane ($\theta \approx 25^{\circ}$), and SLIPS ($\theta \approx 35^{\circ}$), but complete dewetting on the PEGDA hydrogel ($\theta \approx 180^{\circ}$).
    Substrates from left to right: bare glass, PEG-silane (3-[methoxy\-(poly\-ethylene\-oxy)\-propyl]\-tri\-methoxy\-silane), SLIPS of silicone oil (vinyl-terminated polydimethylsiloxane), and chemically cross-linked PEGDA hydrogel (molecular weight 12 kDa).
    Bottom: Schematics of the different surfaces.
    For SLIPS, we created a porous surface by coating the glass with hydrophobic nanoparticles (\textit{black points}) and then infused this surface with silicone oil (\textit{blue}).
    The PEGDA hydrogel was covalently attached to the glass via a bonding layer (\textit{dark red}) and formed by chemically cross-linking the highly hydrated PEG chains (\textit{green}) in the presence of a photoinitiator and irradiation with UV light (not shown). 
    \textbf{(B)} Static contact angle measurements of various droplets in aqueous solution on PEGDA hydrogel (top) versus PEG-silane (bottom).
    From left to right: side view of fluorescently labelled PDDA-STMP coacervate droplets, BSA-PEG droplets, DNA nanostar droplets, and a giant unilamellar vesile (GUV) in solution in the presence of a macromolecular depletant (PEG 100 kDa), resting on a PEGDA 12 kDa hydrogel (top, $\theta \approx 180^{\circ}$) or PEG-silane (bottom, $\theta \approx 130^{\circ}$). 
    For PDDA and STMP droplets and BSA-PEG droplets on PEG-silane, 20\% of the coated glass slides resulted in $\theta \approx 180^{\circ}$, whereas 80\% of slides produced $\theta \approx 130^{\circ}$.
    The contact angle values for the different systems are summarized in Table \ref{tab:pegda_contact_angles}.
    All scale bars are 10 $\mu \mathrm{m}$.
    }
    \label{fig:intro_and_static_theta}
\end{figure*}

To evaluate the performance of the different coatings, we measured the static contact angle, $\theta$,  of protein droplets using 3D confocal fluorescence microscopy (see Experimental Section) \cite{Drelich2020contactangles}. 
Contact angles $0\leq\theta<150^{\circ}$ indicate wetting \cite{Chu2014CSR}, and hence significant interactions between the droplet and the surface. Complete dewetting occurs in the limit of $\theta = 180^{\circ}$, where contact area and interactions between droplet and surface are minimal.
We evaluated wetting performance of different surfaces using droplets of Laf1-AK-Laf1 \cite{faltova2018multifunctional, kuffner2020acceleration}, which is a recombinant fusion protein of the DEAD-box protein Laf1 and adenylate kinase (AK), an enzyme involved in the ATP energy transfer from mitochondria \cite{klepinin2020adenylate}.
The phase separation is driven by weak multi-valent attractions between the intrinsically disordered regions (IDR) of Laf1 \cite{elbaum2015disordered,  Brangwynne2015NatPhys}, resulting in droplets with amphiphillic surfaces. 
We found strong partial wetting of Laf1-AK-Laf1 droplets on bare glass ($\theta \approx 30 ^{\circ}$), PEG-silanized glass ($\theta \approx 25 ^{\circ}$), and SLIPS ($\theta \approx 35 ^{\circ}$) (Fig. \ref{fig:intro_and_static_theta}A). 
In contrast, the PEGDA hydrogel exhibited complete dewetting of Laf1-AK-Laf1 droplets ($\theta \approx 180 ^{\circ}$).

We then tested whether complete dewetting on PEGDA is a unique feature of Laf1-AK-Laf1 droplets by looking at wetting behavior of five other phase-separating systems on the PEGDA hydrogel, using PEG-silanized glass as a control (Fig. \ref{fig:intro_and_static_theta}B). 
We selected the following systems, which spanned a wide range of interface chemistries and driving forces for phase separation:

\paragraph{Complex coacervate} We used poly\-(diallyl\-dimethyl\-am\-monium chloride) (PDDA) and sodium trimetaphosphate (STMP).
This combination of PDDA and a poly\-phosphate salt resembles the PDDA and ATP system, which has been used to study phase-separated micro-compartments \cite{williams2012polymer}.
Complex coacervates are driven by electrostatic attraction of oppositely charged polyions and feature a highly hydrophilic surface \cite{Rumyantsev2021Annurev}.
    
\paragraph{BSA-PEG droplets}
We selected the BSA-PEG system as a globular and segregative complement to the IDR-driven and associative Laf1-AK-Laf1 system.
In this system, BSA and PEG of molecular weight 4 kDa undergo segregative phase separation in the presence of phosphate buffer (100 mM KH$_{2}$PO$_{4}$/K$_{2}$HPO$_{4}$, pH 7) and salt (200 mM KCl) \cite{testa2021sustained}.
Phase separation is driven by PEG-induced depletion interactions between BSA molecules, which themselves are negatively charged at pH 7, resulting in an amphipathic droplet surface.

\paragraph{DNA nanostar droplets} These DNA droplets assemble via specific interactions between complementary base-pairs (hybridization) among so-called DNA nanostars \cite{Biffi2013PNAS}. 
The resultant droplets feature a highly negatively charged surface due to the DNA's phosphate backbone.

\clearpage
    
\paragraph{Giant unilamellar vesicles (GUVs)} We prepared GUVs in the presence of depletion agent (PEG of molecular weight 100 kDa) \cite{spanke2020wrapping, spanke2022dynamics}. GUVs are a model system for biological membranes (phospholipid bilayers) that has been widely used by the scientific community \cite{walde2010giant}. Phospholipid bilayers are 2-dimensional liquids that assemble from amphiphiles (\textit{e.g.} palmitoyl-oleyl-phosphatidylcholine, POPC), and feature a hydrophilic surface that is determined by the lipid headgroups. 
The phosphatidyl choline headgroup of POPC is a zwitterion and carries no net charge.
    
\paragraph{Fluorinated oil droplets} We used FC-70 droplets in water, which have a highly hydrophobic surface. FC-70 was chosen for its relatively high surface tension in water ($\gamma_{FC70}\approx60\pm6~\mathrm{mN \slash m}$, as measured by pendant drop tensiometry), which is roughly $1000 \times$ higher than protein droplets and complex coacervates \cite{ijavi2021surface,Spruijt2010Soft}.

In Figure \ref{fig:intro_and_static_theta}B and Table \ref{tab:pegda_contact_angles} we report the results of this analysis.
Remarkably, we found that all systems showed negligible wetting or adhesion to the PEGDA hydrogel coating, despite significant differences in their chemical properties. 
These results suggest that the PEGDA hydrogel is super-repellent to phase-separated liquids in aqueous buffers and provides an excellent substrate to study protein, DNA, or complex coacervate droplets, as well as phospholipid bilayers.
The coating remains super-repellent in the presence of high molecular weight PEG ($4~\mathrm{kDa}$ and $100~\mathrm{kDa}$).
This suggests that PEG can readily diffuse into or weakly adsorb on the hydrogel because otherwise the depletion effect would induce adhesion of the droplets to the hydrogel surface \cite{Lekkerkerker2011ColloidsAT}.
The PEGDA hydrogel is therefore uniquely suited for studies where PEG is used as macromolecular crowding agent \cite{spanke2020wrapping, spanke2022dynamics,UverskyCrowding2014,WangPEG2007, HanPEG2020}. 
In comparison, droplets partially wetted PEG-silane in most cases (Fig. \ref{fig:intro_and_static_theta}B) with a contact angle of $\theta \approx 130^{\circ}$ for complex coacervates, BSA-PEG protein droplets, GUVs, and FC-70 droplets. 
PEG-silane performed well for DNA droplets ($\theta \approx 180^{\circ}$). 
Occasionally, PEG-silanization yielded complete dewetting of complex coacervates and BSA-PEG protein droplets.
However, in our hands, this performance was achieved in only approximately 20\% of the batches of PEG-silanized substrates.
\begin{table*}
    \centering
    \caption{Approximate values of static contact angle, $\theta$, for different droplets and lipid vesicles systems on PEGDA hydrogel and PEG-silane. For PDDA and STMP droplets and BSA-PEG droplets on PEG-silane, the contact angle is accompanied by the approximate percent of PEG-silanized glass slides that exhibited that contact angle.}
    \vspace{0.2cm}
    \begin{tabular}{c|c|c|c|c|c|c}
         & Laf1-AK-Laf1 & PDDA and STMP & BSA-PEG & DNA Nanostars & POPC & FC-70 \\
         \hline
        PEGDA hydrogel& $180 ^{\circ}$ & $180 ^{\circ}$& $180 ^{\circ}$& $180 ^{\circ}$& $180 ^{\circ}$ & $180 ^{\circ}$\\
        PEG-silane & $25 ^{\circ}$ & $130 ^{\circ}$ to  $180 ^{\circ}$ & $130 ^{\circ}$ to  $180 ^{\circ}$ & $180 ^{\circ}$ & $130 ^{\circ}$ & $130 ^{\circ}$ \\
    \end{tabular}
    \label{tab:pegda_contact_angles}
\end{table*}
\subsection{Dynamic contact angle and droplet sliding}
Sessile droplets exhibit a unique value of contact angle when they have reached thermodynamic equilibrium on a defect-free and non-adapting surface \cite{Butt2018AdaptiveWI}.
This is called the \emph{equilibrium} contact angle.
In most cases, however, the droplet is not in equilibrium due to contact line pinning and the measured $\theta$ deviates from the equilibrium contact angle \cite{Kim2021PRX}.
Furthermore, the measured $\theta$ does not describe dynamics such as droplet sliding on a surface, which involves droplet wetting along its advancing edge and unpinning of the contact line at its receding edge. This is described by two contact angles, $\theta_a$ (advancing) and $\theta_r$ (receding), and their difference, $\Delta\theta$, is known as contact angle hysteresis \cite{shi2018dynamic, eral2013contact}. Contact angle hysteresis is therefore more suitable to evaluate the mobility of a liquid droplet on a  solid substrate. A greater hysteresis generally implies a decrease in droplet mobility and increase in pinning and/or friction with the surface.

The two contact angles are typically measured by depositing a droplet on a flat horizontal substrate and then progressively tilting it \cite{eral2013contact}.
Initially, the droplet will deform under gravity but remain pinned in place by the retention force.
At a critical value of tilting angle $\alpha$, however, the droplet will start to slide down the substrate.
The contact angles measured at the deformed advancing and receding ends of the droplet at the moment of motion onset are $\theta_a$ and $\theta_r$, respectively.
A greater $\Delta\theta$ means greater droplet deformation and thus indicates stronger retention forces.

\begin{figure*}
    \centering
   \includegraphics[width=\textwidth]{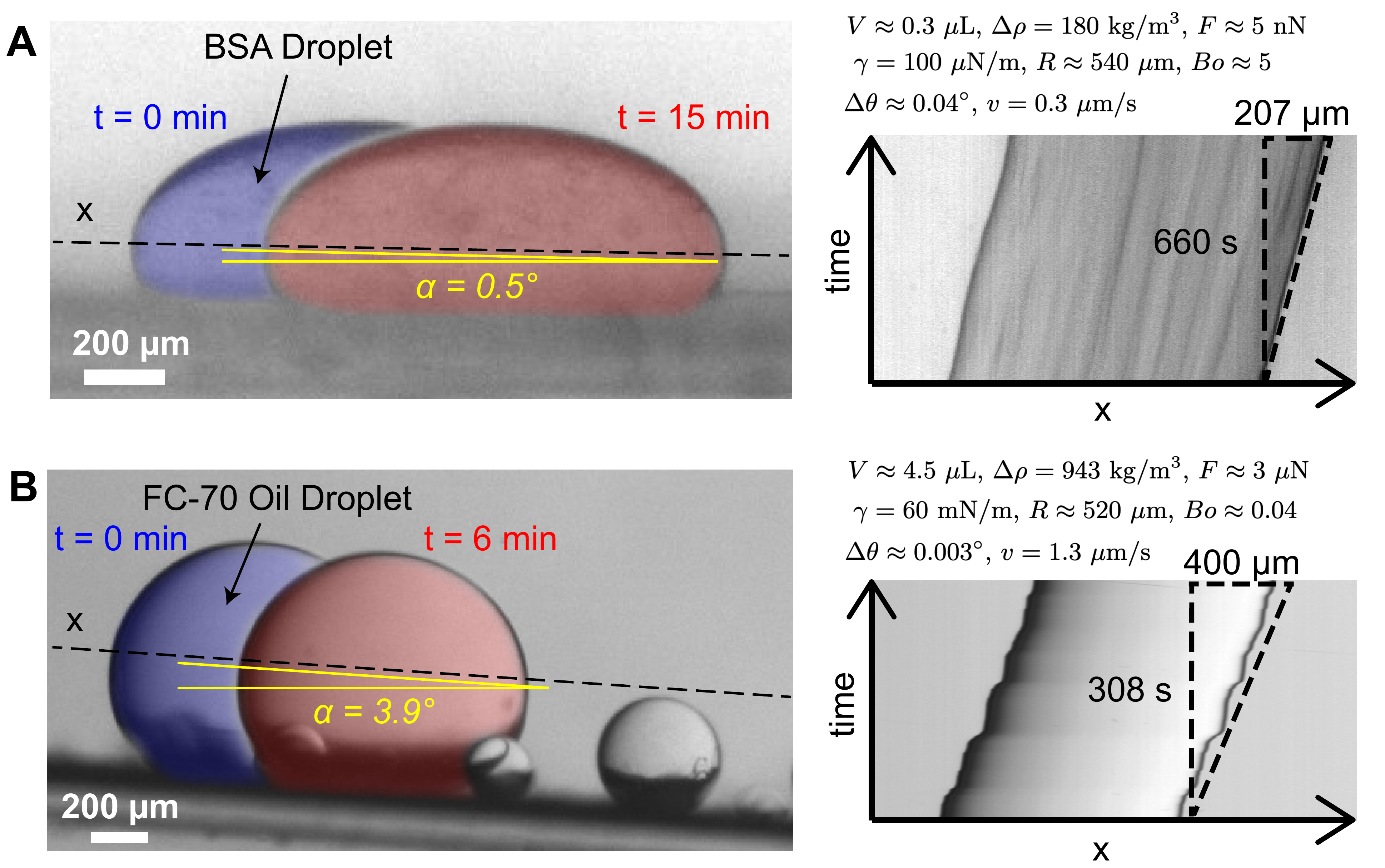}
    \caption{\textit{BSA and FC-70 droplets sliding on a PEGDA 12 kDa hydrogel-coated glass.}
    \textbf{(A)}, Left, overlain images of a BSA droplet sliding on inclined ($\alpha = 0.5^{\circ}$, \textit{yellow}) PEGDA 12 kDa hydrogel, at the beginning ($\mathrm{t}=0~\mathrm{min}$, \textit{blue}) and at the end ($\mathrm{t}=15~\mathrm{min}$, \textit{red}) of the time series.
    Right, kymograph of the BSA droplet with velocity $v=0.3~\mu m/s$.
    \textbf{(B)} Left, overlain images of a FC-70 droplet sliding on inclined ($\alpha = 3.9^{\circ}$) PEGDA 12 kDa hydrogel, at the beginning ($\mathrm{t}=0~\mathrm{min}$, \textit{blue}) and at the end ($\mathrm{t}=6~\mathrm{min}$, \textit{red}) of the time series.
    Right, kymograph of the FC-70 droplet with velocity $v=1.3~\mu m/s$.
    The droplet volume has been estimated considering droplets as ellipsoids that have a circular profile in the x-y plane.
    V: droplet volume, $\mathrm{\Delta \rho}$: density difference between droplet and surrounding solution, $F$ : retention force, $\gamma$: interfacial tension between droplet and surrounding phase, $R$: droplet radius, $Bo$: Bond number, $\Delta\theta$: contact angle hysteresis. All values, except $\Delta\rho$ and $\gamma$, are obtained from analysis of these images. 
    }
    \label{fig:dynamic_ca_and_motion}
\end{figure*}

\begin{figure}
    \centering
   \includegraphics[width=0.8\columnwidth]{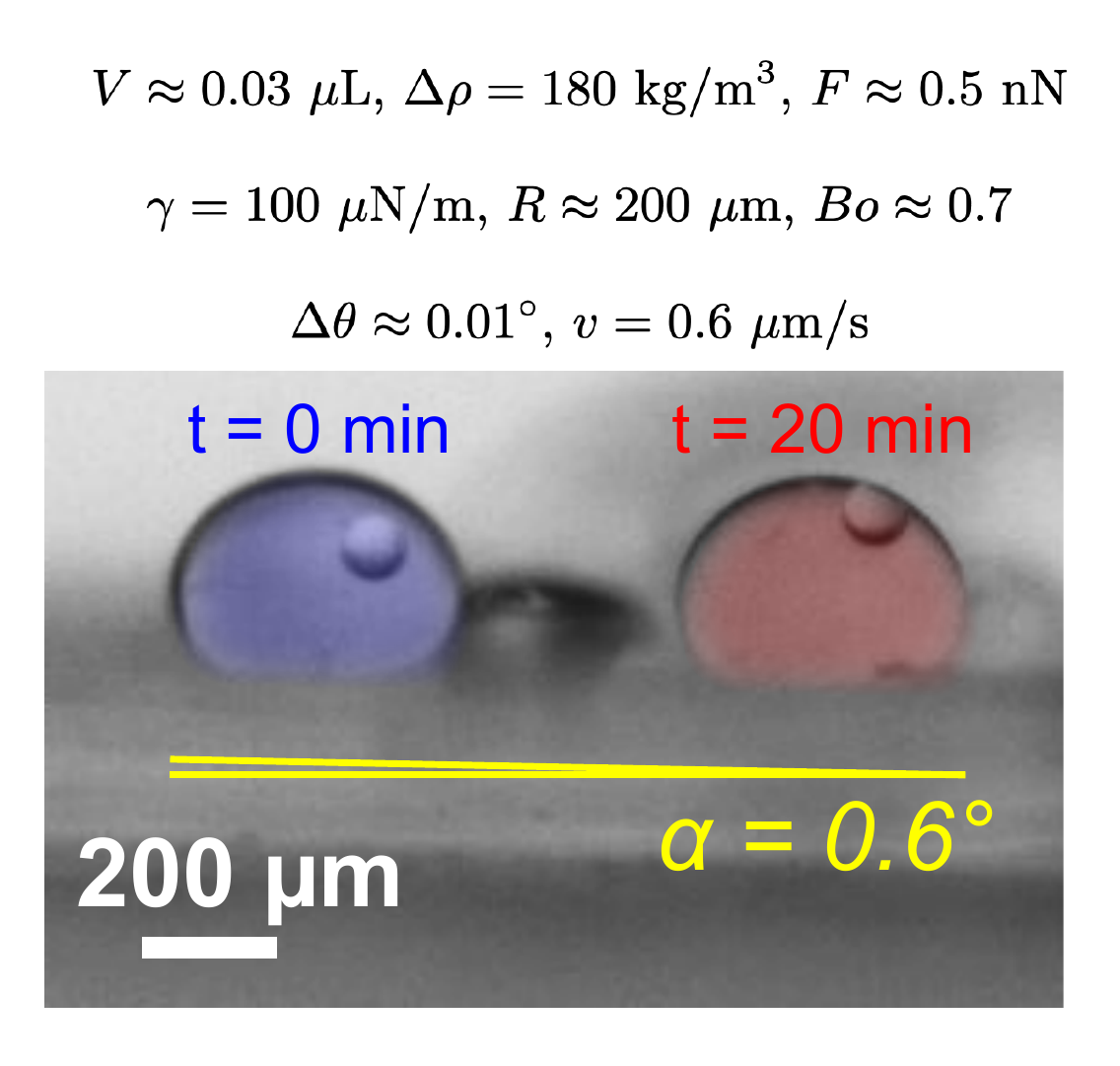}
    \caption{\textit{Supporting Figure: Low critical tilting angle for BSA droplet sliding on PEGDA hydrogel.} 
    BSA droplet of a small volume ($V\approx0.03~\mu\mathrm{L}$) with Bond number $B\approx0.7$ sliding at very low tilting angle ($\alpha=0.6^{\circ}$) and velocity ($v=0.6~\mu\mathrm{m/s}$).
    The corresponding retention force is $F\approx0.5~\mathrm{nN}$ and contact angle hysteresis is $\Delta\theta\approx0.01^\circ$.
    }
    \label{supp_fig:small_BSA}
\end{figure}

First, we measured the critical tilting angles for BSA and FC-70 droplets in solution sliding on the PEGDA hydrogel (Fig. \ref{fig:dynamic_ca_and_motion}).
Inclinations of $\alpha < 1^{\circ}$ were sufficient to make BSA droplets slide on PEGDA hydrogel (Fig. \ref{fig:dynamic_ca_and_motion}A), even for droplets volumes as low as $\approx 0.03 ~ \mu \mathrm{L}$ (Fig. S\ref{supp_fig:small_BSA}).
Similarly, FC-70 droplets started to slide for $\alpha \lessapprox 4^{\circ}$ on the hydrogel (Fig. \ref{fig:dynamic_ca_and_motion}B).
This suggests that minimal pinning occurs for both BSA and FC-70 droplets, despite great differences in their chemical properties.
To our knowledge, the critical tilting angles on the PEGDA hydrogel are smaller than those of the best superomniphobic and superoleophobic coatings reported in literature, which generally require $\alpha$ values of a few degrees and droplet volumes of several microliters for sliding to occur \cite{wong2011bioinspired, chen1999ultrahydrophobic, tang2020lateral, quere2005non}.

We can estimate the retention force, $F$, by equating it with the the gravitational force component parallel to the surface at the onset of tilting, that is, at the critical tilting angle $\alpha$:
\begin{equation}\label{eq:ret_force_eq}
    F = \Delta\rho \cdot g \cdot V\cdot \sin\alpha,
\end{equation}
where $\Delta\rho$ is the density difference between the droplet and the surrounding solution, $g$ is the gravitational acceleration, and $V$ is the droplet volume.
Using Eq. \ref{eq:ret_force_eq} and $\Delta\rho=180~\mathrm{kg/m^{3}}$ \cite{testa2021sustained}, we obtain an estimated retention force $F\approx5~\mathrm{nN}$ for a typical BSA droplet of volume $0.3~\mu\mathrm{L}$ (Fig. \ref{fig:dynamic_ca_and_motion}A), which is at least one order of magnitude lower than the lowest values previously reported for droplets of similar size, including FC-70 droplets on SLIPS \cite{wong2011bioinspired}, water droplets on superhydrophobic surfaces \cite{backholm2020water}, ethylene glycol droplets on silicon wafers \cite{extrand1990retention}, and water droplets on SOCAL \cite{DanielPRL2018}.
We found that BSA droplets with small volumes (\textit{e.g.} $0.03~\mu\mathrm{L}$) could exhibit retention forces as low as $0.5~ \mathrm{nN}$ (Fig. S\ref{supp_fig:small_BSA}).
For a typical FC-70 droplets of volume $4.5~\mu\mathrm{L}$, using $\Delta\rho=943~\mathrm{kg/m^{3}}$, we estimate $F\approx3~\mu\mathrm{N}$  (Fig. \ref{fig:dynamic_ca_and_motion}B), which is comparable to the performance of SLIPS for various droplets in air ($V\approx4.5~\mu\mathrm{L}$, $F\approx1~\mu\mathrm{N}$) \cite{wong2011bioinspired}.

Indeed, our low retention forces are not directly comparable to values reported for different liquids because the retention force is directly proportional to surface tension. 
Instead, droplet pinning is better quantified by $\Delta\theta$, which is independent of surface tension and droplet size.
However, it is challenging to accurately determine $\theta_a$ and $\theta_r$ by visual analysis of our relatively low-resolution images because for such small values of $\alpha$ needed to trigger sliding, the droplets are essentially undeformed at the onset of motion \cite{Liu2019Science, backholm2020water}.
Under these circumstances, one can instead estimate the difference $(\cos \theta_r - \cos \theta_a)$ from the tilting angle $\alpha$ using a force balance between the retention force and gravity \cite{furmidge1962studies, quere2005non,elsherbini2006retention} (see Supporting Information for  details):
\begin{equation}\label{eq:costhetaforce}
    (\cos\theta_r - \cos\theta_a) = \frac{\Delta\rho \cdot g \cdot V\cdot{\pi^3}}{48R\cdot\gamma}\cdot \sin\alpha,
\end{equation}
where, $R$ is the droplet radius as seen from the side and $\gamma$ is the interfacial tension of the droplet to its surrounding phase (surface tension).
This relationship between contact angle hysteresis and the critical tilting angle is best captured by the Bond number, $Bo$, which quantifies the relative strength of gravitation forces, which tend to move (and deform) the droplet, to its surface tension:
\begin{equation}
    Bo = \frac{\Delta\rho\cdot g\cdot R^2}{\gamma}.
\end{equation}
Taylor expanding $\cos\theta$ around $\theta\approx 180^\circ$ and substituting $V\approx\frac{4\pi}{3}R^3$ in Eq. \ref{eq:costhetaforce} yields
\begin{equation}\label{eq:get_hysteresis}
    \Delta\theta \approx \frac{\pi^3}{36}\cdot Bo\cdot \sin \alpha.
\end{equation}
For the droplets in Fig. \ref{fig:dynamic_ca_and_motion}, Eq. \ref{eq:get_hysteresis} yields $\Delta\theta\approx0.04^\circ$ for BSA droplets (using $\gamma_{BSA} = 100 \pm 30~\mu\mathrm{N/m}$ \cite{testa2021sustained}) and $\Delta\theta\approx0.003^\circ$ for FC-70 droplets (using $\gamma_{FC70} = 60 \pm 6~\mathrm{mN/m}$).
The PEGDA hydrogel performs well as a low-adhesion surface, for which the criteria $\Delta\theta < 0.2$ must be satisfied \cite{de2004capillarity}.

\begin{figure*}[h]
    \centering
   \includegraphics[width=\textwidth]{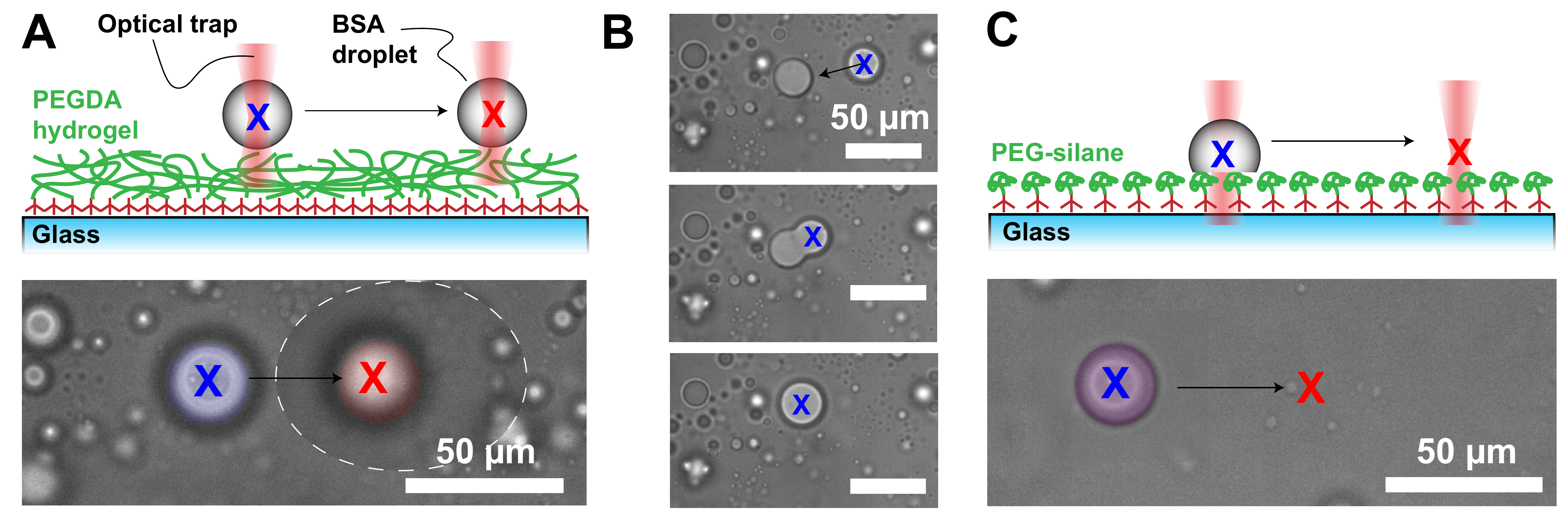}
    \caption{\textit{Optical micromanipulation of BSA droplets in contact with PEGDA hydrogel}
    \textbf{(A)} Top, schematic of micromanipulation of BSA droplets (\textit{gray}) with an optical trap (\textit{red}) on PEGDA hydrogel (\textit{green}). We grabbed a floating droplet from solution gently pushed against the hydrogel (\textit{blue}) and then moved the droplet along the hydrogel surface (\textit{red}) to verify the absence of pinning. Bottom, bright field images corresponding to experimental sequence, where 'X' marks the location of the center of the trap before (\textit{blue}) and after (\textit{red}) movement of the trap. White dashed oval indicates region where image after droplet movement is superimposed.
    \textbf{(B)} Bright field image sequence of a trapped BSA droplet moved to another droplet and their subsequent fusion process.
    \textbf{(C)} Top, schematic of micromanipulation of BSA droplets (\textit{gray}) with an optical trap (\textit{red}) on PEG-silane (\textit{green}). After trapping and lowering the droplet to the surface (\textit{blue}), it partially wetted the PEG-silane, and we could no longer move the droplet (\textit{red}). Bottom, bright field images corresponding to experimental sequence. Original image sequences are found in Fig. S\ref{supp_fig:tweezers}.
    All scale bars are 50 $\mu \mathrm{m}$.
    }
    \label{fig:tweezers}
\end{figure*}

We then tested whether absence of pinning and/or friction extends to the smallest droplets. 
To this end, we used optical tweezers to manipulate microscopic BSA droplets of radius $\sim10~\mu \mathrm{m}$ (Fig. \ref{fig:tweezers}). 
We trapped BSA droplets in solution and gently pushed them onto the coated glass slide surface, and then tried to move the droplet along the surface. 
We were able to freely micro-manipulate BSA droplets, even after they were pushed down onto the PEGDA hydrogel coating with the optical traps (Fig. \ref{fig:tweezers}A), indicating that the dewetting behavior remains stable with respect to mechanical perturbations. 
In particular, we could bring droplets together to induce fusion (Fig. \ref{fig:tweezers}B). 
The fusion process occurred freely for droplets resting on the PEGDA hydrogel. In contrast, BSA droplets brought down to a PEG-silane coating immediately wetted the surface, and thus became immobilized (Fig. \ref{fig:tweezers}C).

Taken together, the micromanipulation experiments, the small retention forces, and the static contact angles of nearly $180^{\circ}$ convincingly demonstrate that protein and oil droplets on the PEGDA 12 kDa hydrogel behave similarly to water droplets on superhydrophobic surfaces under dynamic conditions \cite{quere2005non}.
In other words, interaction between droplets and the PEGDA hydrogel are minimal, and the droplets are free to roll and move with negligible pinning.
\subsection{Substrate morphology and stability}
Hydrogels may experience surface adaption, in which their surface properties change over time \cite{de2004capillarity}. 
This occurs when the gel's polymer chains undergo conformational changes to increase favorable interactions at the surface \textit{e.g.} with a droplet resting on its surface \cite{Holly1975WettabilityOH, Butt2018AdaptiveWI}. 
The appearance of such interactions would introduce adhesion and increase droplet wetting on the surface. 
In principle, such rearrangements could also occur at the surface of a droplet among its constituent macromolecules. 

\begin{figure}
   \includegraphics[width=1\columnwidth,left]{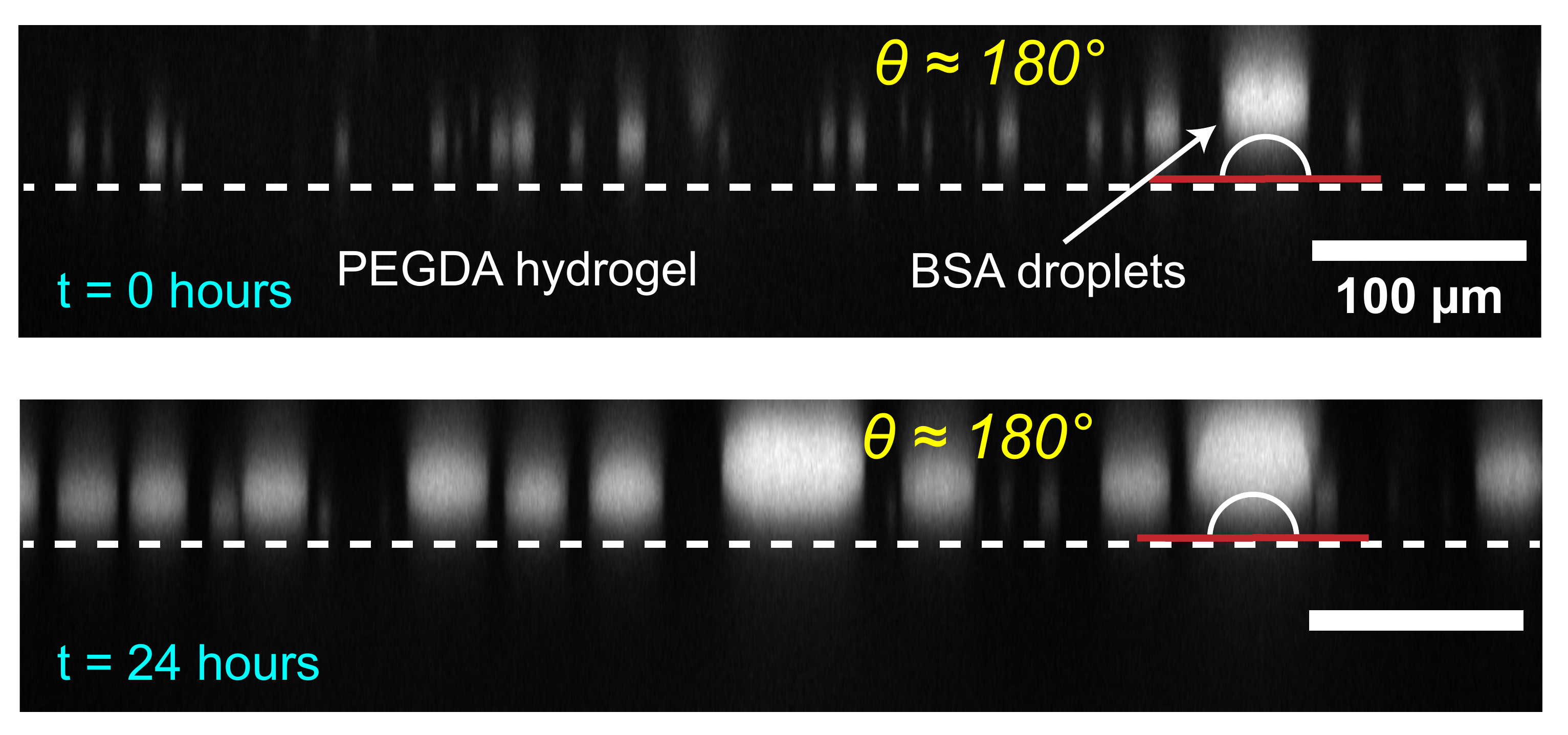}
    \caption{\textit{Coating stability and PEGDA hydrogel.}
    Side view of BSA droplets resting on a PEGDA hydrogel over the course of 24 hours at room temperature. 
    Complete dewetting with a static contact angle $\theta\approx 180^{\circ}$ is retained over the whole time period.
    The gel appears even and without defects over the entire field of view of approximately $1~\mathrm{mm}^2$.
    Scale bars are 100 $\mu \mathrm{m}$.
    }
    \label{fig:morphology_durability}
\end{figure}

We tested the stability of the PEGDA hydrogel and whether surface adaption was taking place by recording static contact angles of BSA droplets over the course of 24 hours and area of $0.7 ~ \mathrm{mm} \times 0.7 ~ \mathrm{mm}$ (Fig. \ref{fig:morphology_durability}). 
We found that the BSA droplets remained dewetted on the PEGDA hydrogel for at least 24 hours. 
These data show that the PEGDA hydrogel possesses sufficient stability for experiments lasting for prolonged time periods, and that surface adaption is negligible on that time scale.

PEG-silane and the SLIPS we used here can be considered flat since they are only a few nanometers thick and thus follow the contour of the underlying glass slide \cite{schroen1995influence}.
PEGDA hydrogels, on the other hand, are prepared by capillary spreading of the PEGDA-photoinitiator solution on the glass slide prior to curing, resulting in a thickness of several micrometers.
They are also reported to undergo modest shrinkage of approximately $7 \%$ during curing \cite{schmidt2015monitoring} and in order to avoid wrinkling, the hydrogel layer had to be sufficiently thin.
Moreover, we cured the hydrogel between two glass slides, one of which needed to be mechanically peeled off before use.
The soft hydrogel would therefore be subjected to substantial stresses that could introduce defects, despite our use of an anti-stick coating.
However, the hydrogel appears flat over the observed field of view of view (Fig. \ref{fig:morphology_durability}). 
The hydrogel thickness is approximately $50 ~ \mu \mathrm{m}$, which closely matches the expected value calculated from the glass slide dimensions ($22 ~ \mathrm{mm} \times 22 ~ \mathrm{mm}$) and the pipetted volume of the PEGDA-photoinitiator solution ($20 ~ \mu \mathrm{L}$).
Furthermore, the droplets appear to sit on the surface of the gel and do not visibly penetrate it.
Areas as large as $1 ~ \mathrm{mm}^2$ are easily found where the performances of the PEGDA gel are optimal, and the morphology of the coating is relatively smooth and even.

\begin{figure*}
   \includegraphics[width=0.8\textwidth]{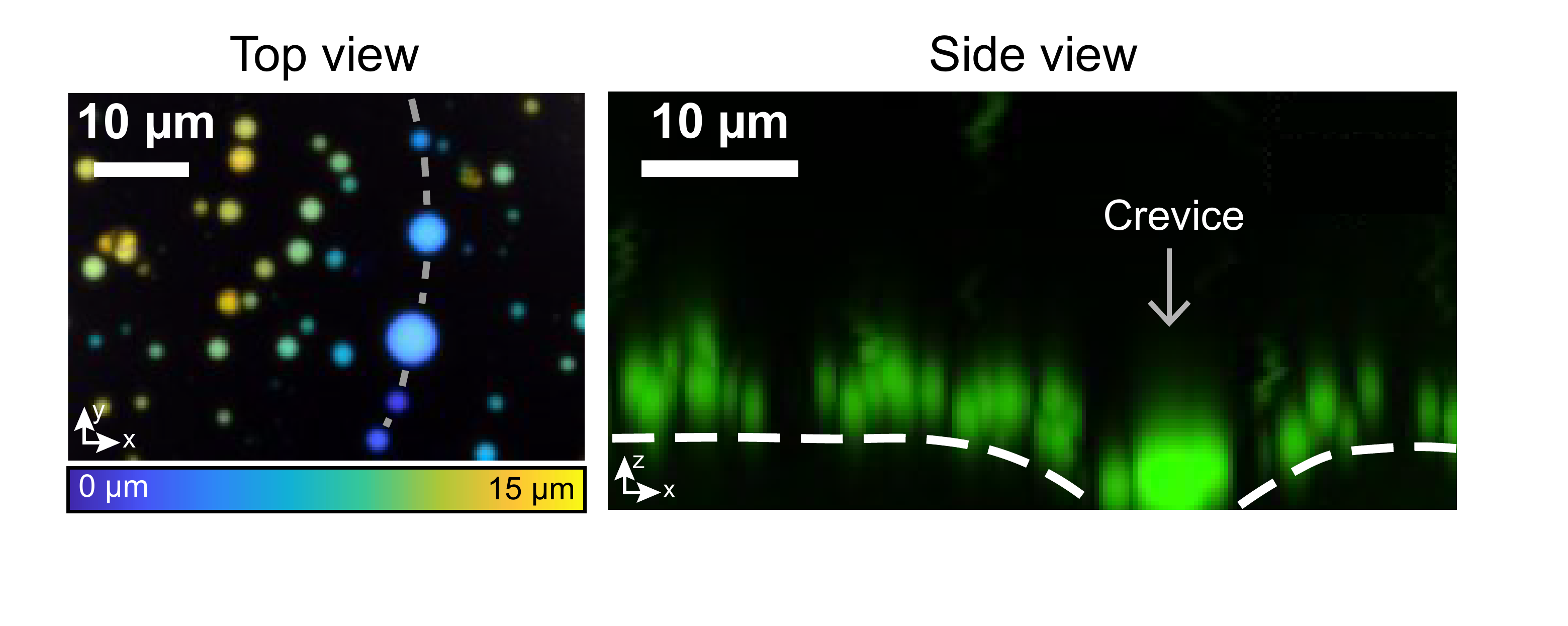}
    \caption{\textit{Supporting Figure: Crevice-like defects on the surface of a PEGDA hydrogel.} 
   Left side: z-projections of top view of an area of the hydrogel, color coded in function of the height from the lowest recorded z-plane. Droplets sitting lower appear blue, while droplets sitting higher appear green and then yellow. The dashed line represents the bottom of the crevice. Right side: side view of the same area of the sample. The dashed line represents the surface of the PEGDA hydrogel. All scale bars are 10 $\mu \mathrm{m}$.
    }
    \label{supp_fig:crevice}
\end{figure*}

During droplet sliding experiments, we found that over areas exceeding $1~\mathrm{mm}^2$, PEGDA hydrogels can be uneven and have a few defects.
While this is not particularly problematic for experiments with micron-sized droplets, as areas of the coating with good performance are typically larger than the field of view, it already becomes relevant for droplet sizes and trajectory lengths we observed for the dynamic contact angle measurements.
The uneven kymograph accompanying the sliding FC-70 droplet in Fig. \ref{fig:dynamic_ca_and_motion}B also suggests that the PEGDA hydrogel contains some unevenness or minor defects that weakly pin the oil droplet with separations $\sim100~\mu\mathrm{m}$.
These defects were neither apparent in Fig. \ref{fig:morphology_durability} nor did they affect the sliding of BSA droplets (Fig. \ref{fig:dynamic_ca_and_motion}).
However, in a few cases, we found visible defects in the form of crevices with depths of several micrometers, as shown in Fig. S\ref{supp_fig:crevice}, where the imperfections are visualized by the presence of numerous small droplets on the hydrogel surface.
We attribute these imperfections to damages sustained by the gel during curing or by the excessive deformations induced in the hydrogel upon peeling of the top glass slide.
We anticipate that the gel's robustness can be significantly improved by optimizing the polymer content, amount of cross-linker, and curing time.

\subsection{Conditions for dewetting}
The remarkable consistency of dewetting across a variety of chemically distinct droplet systems and the lack of surface adaption over time 
strongly suggests that the mechanism is independent of the droplet's detailed chemical properties. 
To rationalize this robust behavior, we adopt a simplified picture of a hydrogel as a porous material with polymer volume fraction, $\phi$.
This allows us to use the Cassie-Baxter equation to express the \textit{apparent} contact angle $\theta_{app}$ of a droplet resting on the hydrogel \cite{Law2016SurfaceWC, Cassie1944WettabilityOP}:
\begin{equation}
    \label{eq:CB}
    \cos{\theta_{app}} = \phi( 1 + \cos{\theta} ) -1,
\end{equation}
where $\theta$ is the droplet's contact angle with a surface comprising pure polymer ($\phi \sim 1$).
Dewetting requires $\cos(\theta_{app}) \rightarrow -1$, which can readily be satisfied for $\theta > 90^{\circ}$ and $\phi$ sufficiently small.
For highly swollen hydrogels ($\phi\ll1$), complete dewetting should occur when the polymer has even a slight preference for buffer over droplet constituents.

Equation \ref{eq:CB} suggests that most swollen hydrogels will create a super-repellent surface for droplets in aqueous buffer.
However, crosslinked polyethylene glycol (PEG) moieties are particularly suitable because they are neutrally charged and extremely hydrophilic.
Instead of interacting with hydrophilic molecules (\textit{e.g.} polar side-chains) or sticking to hydrophobic molecules (\textit{e.g.} fluorinated oil), PEG prefers to be hydrated by water.
In addition, PEG does not engage in electrostatic interactions with charged polymers because it is neutral.
As a result, PEG doesn't stick to most proteins, nucleic acids, complex coacervates, or lipid membranes.
\section{Conclusion}
We developed a novel surface coating that repels droplets in aqueous solution while being optically transparent and thus ideally suited for light microscopy. 
Our coating is based on a chemically cross-linked PEGDA hydrogel that is covalently bound to a microscope glass slide.
Complete dewetting on the PEGDA hydrogel occured for droplets with remarkably diverse chemical properties including depletion-induced BSA droplets, IDR-containing protein droplets, complex coacervates, DNA nanostar droplets, fluorinated oil droplets, as well as phospholipid bilayers in the presence of a macromolecular crowding agent, PEG.
All systems exhibited highly spherical shapes with static contact angles $\theta\approx180^{\circ}$.
In addition, we found minimal pinning and/or friction of BSA droplets and FC-70 droplets on the PEGDA hydrogel.

PEGDA hydrogels thereby offer a long sought-after surface treatment with vanishing adhesion to a wide range of aqueous dispersions. 
More robust and scalable coatings of this kind could find applications as anti-fouling surfaces for biomedical and environmental applications.
\clearpage
\section{Experimental Section}
\label{sec:methods}
\subsection{PEGDA 12 kDa synthesis}
PEGDA can be synthesized by starting from PEG at the desired molecular weight and acryolyl chloride \cite{delong2005covalent}.
The objective of the synthesis is to attach an acrylate group from the acryolyl chloride molecule to each end of the polyethylene glycol.
While PEGDA can be easily found commercially at long molecular weights, its price can be high and it often contains curing inhibitors to increase its shelf life, whose complete removal may be challenging.
A custom-made synthesis allows easier selection of the final molecular weight and yields a product that is more pure. 

We started by weighing $10 ~ \mathrm{g}$ of PEG $12 ~ \mathrm{kDa}$ (Alfa Aesar, 042635-30)  and added to a Schlenk flask with a magnetic stirrer.
Oxygen interferes with the first step of the synthesis and significantly reduces the yield.
We therefore removed the oxygen by repeating four times a cycle of air removal with a vacuum pump followed by a nitrogen purge.
Next, we injected in the flask $68 ~ \mathrm{mL}$ of dry dichloromethane (DCM) (Acros Organics, 34846) and $0.27 ~ \mathrm{mL}$ of acryolyl chloride, yielding a 1:1 molar ratio of acroyl chloride to the hydroxy endgroups on the PEGDA.
The reaction between PEGDA and acroyl chloride is reversible.
We therefore added $0.46 ~ \mathrm{mL}$ of triethylamine (VWR, 28745.296) (at 2:1 excess to the acroyl chloride) to quench the reverse reaction.
In all cases, we made sure to avoid contact between solutions and oxygen.
We then covered the flask with aluminium foil and left it stirring overnight.

The next day, we transferred the contents of the Schlenk flask to a round-bottom flask and evaporated roughly half of the DCM with a Rotovap (Rotovapor R-300, Büchi) at a temperature of $50 ~ ^{\circ} \mathrm{C}$ and pressure of $690$ $\mathrm{mbar}$.
We then prepared roughly $300 ~ \mathrm{mL}$ of cold diethyl ether in a beaker by placing the beaker into an ice bath. The solution from the Rotovap was then transferred to the cold diethyl ether in a dropwise manner using a pipette.
A white solid immediately precipitated.
The liquid was subsequently vacuum-filtered and the powder remaining on the filter was further washed three times with cold diethyl ether.
The powder was then transferred to a Petri dish and allowed to dry under vacuum for a couple days.
\subsection{PEGDA 12 kDa hydrogel substrate preparation}
\begin{figure*}
    \centering
   \includegraphics[width=0.9\textwidth]{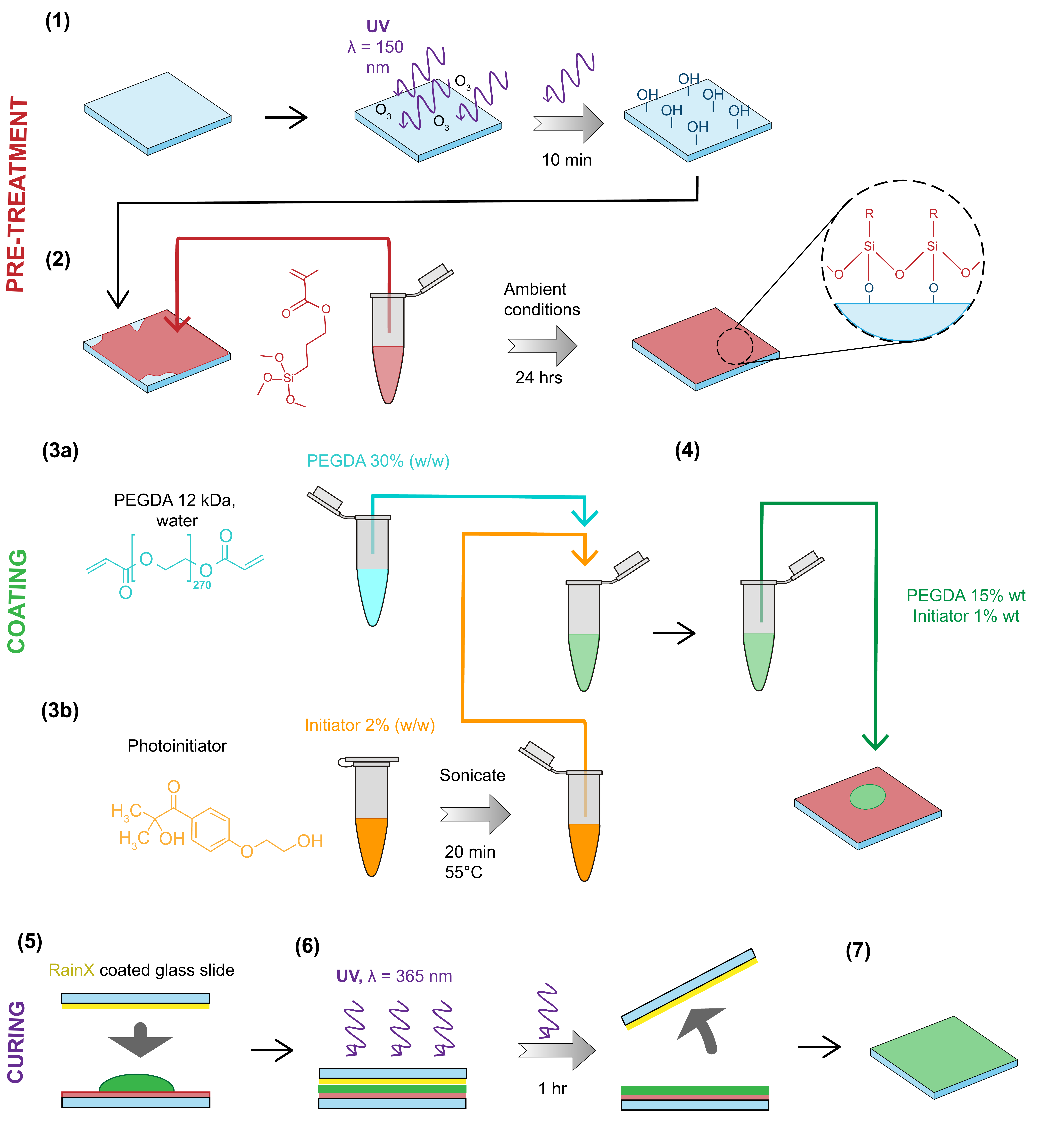}
    \caption{\textit{PEGDA $12 ~ \mathrm{kDa}$ substrate preparation} 
    \textbf{Pre-treatment} \textbf{(1)} Clean glass slides are treated with UV-Ozone for 10 minutes and then \textbf{(2)} coated with a solution containing 3- (tri\-ethoxy\-silyl) propyl\-methacry\-late (silane coupling agent, \textit{red}).
    The reaction is allowed to complete over the course of 24 hours.
    \textbf{Coating} \textbf{(3a)} A solution of PEGDA 30 $\%$ by weight in water is made (\textit{cyan}). \textbf{(3b)} At the same time, a solution of photoinitiator 2 $\%$ by weight in water is made (\textit{orange}) and sonicated for 20 minutes at $55^\circ C$ to assist solubilization. The PEGDA and photoinitiator solutions are then mixed together at 1:1 volume ratio (\textit{green}) to yield final concentrations of $15\%~\mathrm{w/w}$ PEGDA and $1\%~\mathrm{w/w}$ photoinitiator and then \textbf{(4)} added to a pre-treated glass slide.
    \textbf{Curing} \textbf{(5)} The drop of the mixture is sandwiched between the pre-treated glass slide and a glass slide coated with RainX\textsuperscript{\textregistered} (\textit{yellow}) and then \textbf{(6)} cured under UV light for 1 hour. 
    \textbf{(7)} The top glass slide is then removed to expose the cured PEGDA hydrogel.
    }
    \label{fig:pegda_substrate_making}
\end{figure*}

We pre-treated the glass slides to allow chemical bonding of the PEGDA layer to the glass \cite{yuk2016tough}.
We took $22 ~\mathrm{mm}\times 22 ~ \mathrm{mm} ~ \#1.5$ thickness glass slides (VWR) and washed them once with MilliQ water, then ethanol, and again with MilliQ water, followed by drying with an air gun.
We then treated the glass slides with an UV-Ozone cleaner (Bio-force Nanosciences, Pro-Cleaner) for at least 10 mintues, in order to allow the formation of reactive $\mathrm{OH}$ bonds on the glass surface (Figure \ref{fig:pegda_substrate_making} (1)).
During UV-ozone cleaning, we prepared a solution of $900~\mu\mathrm{L}$ analytical grade ethanol (Fisher Chemicals) and $50~\mu\mathrm{L}$ MilliQ water.
We then added $3~\mu\mathrm{L}$ 3-(tri\-me\-thoxy\-silyl) pro\-pyl\-me\-tha\-cry\-la\-te (Tokyo Chemical Industry, M0725) to the ethanol-water mixture, followed by brief vortexing.
This compound is a silane coupling agent that consists of a silicon-containing moiety attached to an acrylate group.
The silicon-containing end can react with $\mathrm{OH}$ groups on the glass surface to create a silicate bond and displace a water molecule, while the acrylate end can react with the crosslinker and get directly integrated into the PEGDA network.
The solution was kept at room temperature on the bench for at least 5 minutes prior to use.

Within 5 minutes of removing the glass slides from the UV-Ozone cleaner, we added $30~\mu\mathrm{L}$ of the silane coupling agent to each slide.
The solution was spread over the glass slides by moving each slide in different directions to ensure even distribution of solution.
The reaction was allowed to proceed for 3 minutes on the bench, after which it was quenched by placing the glass slides into a Petri dish containing analytical grade ethanol (Figure \ref{fig:pegda_substrate_making} (2)).
The glass slides were then gently wiped with a Kimwipe wetted with analytical grade ethanol and subsequently dried with an air gun.

The next day, we prepared in a glass vial a $2~\mathrm{mL}$ solution of photoinitiator 2-Hy\-dro\-xy-4’-(2-hy\-dro\-xy\-etho\-xy)-2-me\-thyl\-pro\-pio\-phe\-no\-ne, also known as Irgacure 2959 (Merck, 410896), at a final concentration of approximately $2  \% ~ \mathrm{(w/w)}$.
We wrapped the vial in aluminium foil to avoid exposure to light and sonicated it in a bath sonicator for 20 minutes at $55 ~ ^{\circ} \mathrm{C}$ to allow its complete dissolution.
The solution was subsequently allowed to cool to room temperature.
During sonication, we prepared $600~\mu\mathrm{L}$ of PEGDA $12 ~ \mathrm{kDa}$ stock at $30\% ~ \mathrm{(w/w)}$ in MilliQ water (Figure \ref{fig:pegda_substrate_making} (3)).
We found that the smoothness of the final hydrogel increased with increasing PEGDA concentrations, and we found that the solubility limit of PEGDA $12 ~ \mathrm{kDa}$ in water is about PEGDA $40\% ~ \mathrm{(w/w)}$.
We then mixed the initiator solution 1:1 with the PEGDA $12 ~ \mathrm{kDa}$ ($600~\mu\mathrm{L}$ each, with final concentrations $1\% ~ \mathrm{(w/w)}$ of photoinitiator and $15\% ~ \mathrm{(w/w)}$ of PEGDA $12 ~ \mathrm{kDa}$.

We also prepared a second set of glass slides to cover the glass slides treated with PEGDA during UV curing. 
These new slides were, like before,  washed with water, ethanol, water, and then dried with an air gun.
We then rubbed these slides with a Kimwipe drenched in RainX\textsuperscript{\textregistered} Original Glass Water Repellent to create a non-stick coating, and left these slides out to dry on the bench for a few minutes. 
For UV curing of the hydrogel, we added around 20 $\mu \mathrm{L}$ of the 1:1 PEGDA-photoinitiator solution to each pre-treated glass slide (Figure \ref{fig:pegda_substrate_making} (4)) and then covered each one with a  RainX\textsuperscript{\textregistered}-treated glass slide (Figure \ref{fig:pegda_substrate_making} (5)). In this way, a thin layer of the PEGDA-initiator mixture was allowed to spread via capillary action between the two glass slides thereby creating a uniform coating.
The resultant glass slide sandwiches were put under a UV lamp at wavelength $\lambda = 365 \mathrm{~nm}$ and power $15~\textrm{Watts}$ (Analytik Jena, UVP XX 15BLB) for 1.5 hours to allow complete curing of the PEGDA gel (Figure \ref{fig:pegda_substrate_making} (6)).
The cured slides were stored in a closed Petri dish submerged in MilliQ water until further use.

A glass sandwich was only opened immediately before experiment (Figure \ref{fig:pegda_substrate_making} (7)).
Importantly, due to the permeability of the hydrogel to small solutes, it had to be equilibrated by soaking it for 30 minutes with  supernatant stemming from the droplet suspension.
After equilibration, the droplets suspension was added to the PEGDA hydrogel and the system was sealed in order to avoid evaporation.
To this end, we partially removed a portion of the PEGDA hydrogel using a razor to provide adhesive surface for silicone isolators (Press-to-Seal\textsuperscript{\texttrademark}, P24744) and then covered the sample with a cover glass.
Under these conditions, the droplets and hydrogel could remain stable for prolonged time periods $\geq24~\mathrm{h}$.
\subsection{PEG-silane substrate preparation}
First, we prepared the PEG-silanization solution by mixing $500~\mathrm{mL}$ toluene (Merck, 32249) and $2.3~\mathrm{mL}$ 2-[Methoxy\-(poly\-ethylene\-oxy)\-propyl]\-tri\-methoxy\-silane, 90\%, 6\--9 PEG units (ABCR GmbH, AB111226) in a glass bottle followed by vigorous mixing.
We then added $800~\mu\mathrm{L}$ of aqueous HCl (37\%) (VWR, 12463) to the solution.
After mixing thoroughly, the solution was carefully poured into a large crystallization dish placed under the fume hood while vigorously stirring with a magnetic stirrer.

Next, we washed microscope slides with water, ethanol, and water, followed by drying with an air gun, as described in the PEGDA hydrogel procedure.
As before, the clean glass slides were treated for at least 10 minutes with UV-Ozone.
The UV Ozone-treated glass slides were gently placed into the crystallization dish, making sure that the solution covers the coverslips and that the stir bar does not disturb the slides.
If available, a suitable glass rack may be used at this stage to immobilize and protect the glass slides inside the solution.
The crystallization dish was covered with parafilm to prevent evaporation of toluene and stirred in the fume hood at room temperature for 18 hours. 
Afterwards, the glass slides were rinsed once in toluene, and then twice more with analytical grade ethanol, before being dried with an air gun. 
Finally, the glass slides were stored in a dry chamber until required for use.
\subsection{SLIPS}
Microscope slides were washed with water, ethanol, and water, followed by drying with an air gun, as described above.
Following a previously described procedure \cite{Bradley2019PRL}, we then sprayed silica nanoparticles (Glaco Mirror Coat Zero\textsuperscript{\textregistered}) onto the glass slides and waited 30 minutes to allow the alcohols evaporate. 
The spraying and drying was repeated two more times, but for the last time, we let the coated glass slides dry for 24 hours at room temperature, followed by 1 hour under vacuum. 
The final step to preparing the SLIPS surface was to add a drop of vinyl-terminated polydimethylsiloxane of viscosity 500 cSt (Gelest, DMS-V25) and spin coat for 30 minutes at 1000 RPM.
\subsection{Droplet preparation}
\paragraph{Production of recombinant Laf1-AK-Laf1} Laf1-AK-Laf1 was purified as previously described \cite{kuffner2020acceleration}. Shortly, the plasmid for recombinant expression was codon optimized for \textit{E. coli}, synthetized and cloned into a pET-15b vector by Genewiz (NJ, US). We fused \textit{E. coli} adenylate kinase (AK) with the N-terminal LCD from Laf1 from \textit{Caenorhabditis elegans} (AA 1-168). \textit{E. coli} BL21-GOLD (DE3) cells were used for recombinant expression. Recombinant expression was induced at OD 0.7 with 0.5 mM isopropyl D-thiogalactopyranoside (99\%, PanReac AppliChem) and grown for an additional 16 h at 37 °C. Cells were lysed by 15 x 60 s sonication pulses on ice with 120 s cooling breaks. Cells were lysed in pH 7.5, 50 mM Tris-HCl, 10 mM Imidazol (Sigma, Switzerland), 2 mM 2-beta mercaptoethanol (Sigma, Switzerland) and 500 mM NaCl (all reagents were obtained from Sigma, Switzerland). Laf1-AK-Laf1 was purified by immobilized metal ion affinity chromatography (Chelating Sepharose, GE Healthcare) according to a standard protocol. The protein was further purified by size exclusion chromatography using a Superdex 75 16/600 column (GE Healthcare) assembled on an ÄKTA Prime system (GE Healthcare) using 50 mM Tris at pH 7.5 and 500 mM NaCl as eluent buffer. Final purity of the proteins was assessed by SDS-PAGE electrophoresis. Protein stocks were concentrated to $\sim1000~\mu\mathrm{M}$ and aliquots ($20~\mu\mathrm{L}$) were frozen and stored at $-20^\circ C$ until use. To initiate condensate formation the stock solution was diluted to a final concentration of $20~\mu\mathrm{M}$ in a Tris-HCl pH 8.0 buffer without salt.

\paragraph{Complex coacervate} We made a solution of $20~\mathrm{mM}$ poly\-(diallyl\-dimethyl\-am\-monium chloride), 35\% solution, (Merck, 522376), $6.7~\mathrm{mM}$ sodium trimetaphosphate (Merck, T5508), and $50~\mu\mathrm{M}$ Alexa Fluor\texttrademark\,647 (ThermoFisher, A20006) in MilliQ water.
The concentrations of the poly-cation and poly-anion were chosen to yield a 1:1 molar ratio of positively and negatively charged moieties, respectively.
The dye was added after the PDDA but before the STMP.
    
\paragraph{BSA-PEG droplets}
BSA-PEG droplets were prepared as previously described \cite{testa2021sustained}. Briefly, we prepared a solution of 100 mM potassium phosphate, pH 7, (Acros Organics, 42420 and Merck, 60349), 200 mM KCl (Merck, 60128), $30~\mathrm{g/L}$ bovine serum albumin (BSA) (Merck, A7638) and  $230~\mathrm{g/L}$ PEG 4000 (Alfa Aesar, A16151) in MilliQ water. 
In order to visualize the droplets with fluorescence imaging, a small fraction of the BSA was fluorescently labelled using Alexa Fluor\texttrademark ~594 NHS ester (ThermoFisher, A20004).

\paragraph{DNA nanostar droplets}  The DNA droplets are composed of nanostars, each of which consist of 4 double-stranded DNA oligomers (Integrated DNA Technologies) that hybridize to form a cross-shape. The nanostars have short sticky palindromic overhangs on each arm, which provide weak specific attractions between nanostars with the same overhangs and thus drive droplet formation. 

DNA nanostar droplets were prepared as previously described \cite{Jeon2020JPCB}. 
Briefly, we first generated the DNA nanostars by mixing together the four oligomers, each at $50~\mu\mathrm{M}$ in 10 mM Tris-HCl (pH 7.5), heated to $95^\circ \mathrm{C}$, and then annealed by cooling to $4^\circ C$ at $-0.5^\circ \mathrm{C/s}$.
We created two species of nanostars that do not bind to each other due to incompatible overhang sequences, and a third species, which is a mix of the two overhang sequences and can thus bind both species. 
To visualize the nanostars, we added a small fraction ($1~\mathrm{mol}\%$) of oligomers carrying a covalent fluorescent label, Cy3 and fluorescein (Integrated DNA Technologies) for the first and second species, respectively.
We generated droplets by mixing together the nanostar stock solutions (1:1:0.5 molar ratios between first, second, and third species, respectively) and diluting to $1~\mu \mathrm{M}$ in 10 mM Tris-HCl (pH 7.5) and 500 mM NaCl.
We then heated the samples to $50^\circ\mathrm{C}$ for 30 minutes and cooled to room temperature.
Imaging was performed after 1 hour.
    
\paragraph{Fluorinated oil droplets} We prepared 3M\texttrademark~Fluoriniert\texttrademark~ FC-70 droplets in 1 mL MilliQ water by adding $5~\mu\mathrm{L}$ of FC-70 and $1~\mu\mathrm{L}$ of 1 mg/mL Rhodamine G stock solution in ethanol.
The solution was then vortexed and $20~\mu\mathrm{L}$ of the suspension were immediately transferred into the imaging chamber.
The confocal fluorescence images were inverted to better visualize the dorplets because the fluorescent dye partitioned almost completely into the aqueous phase. 
\subsection{GUV preparation}
We made the GUVS with the electroformation technique \cite{Angelova,Angelova02}. We first prepared a $1~\mathrm{mM}$ solution of lipids, with lipid composition of $99~\mathrm{mol}\%$  1-Palmitoyl-2-oleoyl-sn-glycero-3-phosphocholine (POPC) (Avanti Polar Lipids, 850457C) and $1~\mathrm{mol}\%$ of 1,2-dioleoyl-sn-glycero-3- phosphoethanolamine-N-(lissamine rhodamine B sulfonyl) (ammonium salt) (Rh-DOPE) (Avanti Polar Lipids, 810150C). $50~\mu \mathrm{L}$ of this solution  was deposited on an ITO-plate using a glass syringe (Hamilton). A PDMS-Spacer was placed on the ITO-plate and a second ITO -plate is put on top. The chamber was filled with a solution of $280~\mathrm{mOsm/kg}$ sucrose (BioXtra, $\geq99.5 \%$) (Merck, S7903) and sealed.

The ITO-plates are electrically connected to a signal generator (Keysight 33210A). The electroformation protocol consists of gentle increase in AC-voltage over the course of $25~\mathrm{min}$ from $0$ to $5~\mathrm{V}$ at a fixed frequency of $10~\mathrm{Hz}$. After the voltage reaches $5~\mathrm{V}$, it is left for two hours. Then, keeping the voltage at $5~\mathrm{V}$, the frequency is decreased to $5~\mathrm{Hz}$ and left for another $30~\mathrm{min}$. The specific voltages, frequences, and treatment times are listed in Table S\ref{supp_tab:electroformation}. The chamber was then disassembled and the solution containing vesicles with varying sizes between a few $\mu \mathrm{m}$ and up to $50~\mu \mathrm{m}$ were transferred to a glass vial and stored up to a few weeks.
To prepare GUVs for fluorescence microscopy, 10 $\mu\mathrm{L}$ of the GUV solution was added to $\sim100~\mu\mathrm{L}$ of depletion medium, which consisted of 10 mM NaCl (VWR, 27810), 290 mOsm/kg D-(+)-glucose (BioXtra, $\geq99.5 \%$) (Merck, G7528) and 0.68 \%(w/w) PEG 100 kDa (Merck, 181986) and then transferred to the sample chamber for imaging.
\subsection{3D confocal imaging and contact angle measurement}
We obtained z-stacks of fluorescently labelled droplets resting on substrates (Fig. \ref{tab:pegda_contact_angles}) with a microscope (Nikon, Eclipse Ti2) equipped with a spinning disk confocal system (Yokagawa CSU-X1) using a $40\times$ oil immersion objective lens (Nikon, MRH01401) with a $ 0.3~\mu \mathrm{m}$ step size along the z-axis. 
Images with larger field of view (Fig. \ref{fig:morphology_durability}) were acquired using a $20\times$ air immersion objective lens (Nikon, MRH08230) with a step size of $2.5~\mu \mathrm{m}$.
We processesd the z-stacks with the ImageJ software to derive orthogonal views of the droplets \cite{ImageJ}. 
For the large field of view, we additionally performed a maximum intensity projection along one place in order to display more droplets in the figure. 
We corrected for optical aberrations along the z-axis, which appeared due to the refractive index mismatch between the aqueous sample solution and the objective media \cite{hell1993aberrations, sheppard1997effects}. 
Specifically, we multiplied the nominal z-step size with the factor $k = 0.85$ for the immersion oil and $k = 1.33$ for air.
We used ImageJ to measure the contact angles of the corrected images.
\subsection{Sliding contact angle measurements}
The droplets were imaged using an optical setup with a white $3.5"\times6"$ LED backlight (Edmund Optics), a $0.5 - 1.0\times$VariMagTL\texttrademark ~~Telecentric Lens (Edmund Optics), and a CMOS Camera from Thorlabs (DCC3240M – High-Sensitivity USB 3.0, $1280 \times 1024$, Global Shutter, Monochrome Sensor). 
The use of a telecentric lens is important to ensure accurate size measurements. Samples were held in conventional polystyrene spectrophotometry cuvettes, in which a piece of the coated glass was placed. The cuvette was filled with the continuous phase (supernatant), into which droplets of the co-existing dispersed phase were transfered using a micropipette. Tilting angles were adjusted using a Thorlabs goniometer.
\subsection{Droplet micromanipulation with Optical tweezers}
Experiments using optical tweezers were done with Nikon Ti Eclipse inverted microscopes using a $60\times$ water immersion objective lens. The trapping laser (Omicron Laserage Laserprodukte GmbH, LUXX 785-200 Laser) had a wavelength of 785 nm and a maximum power of 200 mW. Images were captured with a Hamamatsu ORCA-Flash 4.0, C13440. The laser output was vertically polarized and sent through a half-wave plate and a polarizing beamsplitter. The laser power at the sample and thus the trap strength was controlled by appropriately orienting the half-wave plate. After the beamsplitter, the light was given circular polarization using a quarter-wave plate.
\subsection{Measurement of surface tension of FC-70 in water}
We measured the interfacial tension of FC-70 droplets in water using a home-made pendant-droplet tensiometry setup. 
We suspended droplets from a G18 needle into a cuvette filled with MilliQ water. 
We imaged the stable suspended droplets with the same optical setup as for sliding contact angle measurements and analyzed them with axisymmetric drop shape analysis \cite{Ro1997AxisymmetricDS}.
We used $\rho_{H_2O}=997~\mathrm{kg/m^3}$ and $\rho_{FC70}=1940~\mathrm{kg/m^3}$, \textit{i.e.} $\Delta\rho=943~\mathrm{kg/m^3}$.
\section{Statistics and Reproducibility}
No statistical method was used to predetermine sample size. No data were excluded from the analyses. The experiments were not randomized. The Investigators were not blinded to allocation during experiments and outcome assessment.
\section{Data  Availability}
The datasets generated during and/or analysed during the current study are available from the corresponding author on reasonable request.
\section{Acknowledgments}
This work was supported by Ambizione Grant 202214 from the Swiss National Science Foundation to AAR.
We thank Dominic Gerber and Yanxia Feng for insightful discussions.
\section{Author Contributions Statement}
AAR, AT, RWS, and ERD wrote the paper. AAR, AT, HTS, MY, EJP, PA, RWS, and ERD designed the experiments. AAR, AT, HTS, MY, EJP, AMK, and MY performed the experiments. AAR, AT, HTS, EJP, RWS, and ERD performed the data analysis.
\section{Competing Interests Statement}
The authors have no competing interests.
%

\begin{thebibliography}{79}
\expandafter\ifx\csname natexlab\endcsname\relax\def\natexlab#1{#1}\fi
\expandafter\ifx\csname bibnamefont\endcsname\relax
  \def\bibnamefont#1{#1}\fi
\expandafter\ifx\csname bibfnamefont\endcsname\relax
  \def\bibfnamefont#1{#1}\fi
\expandafter\ifx\csname citenamefont\endcsname\relax
  \def\citenamefont#1{#1}\fi
\expandafter\ifx\csname url\endcsname\relax
  \def\url#1{\texttt{#1}}\fi
\expandafter\ifx\csname urlprefix\endcsname\relax\def\urlprefix{URL }\fi
\providecommand{\bibinfo}[2]{#2}
\providecommand{\eprint}[2][]{\url{#2}}

\bibitem[{\citenamefont{Banani et~al.}(2017)\citenamefont{Banani, Lee, Hyman,
  and Rosen}}]{banani2017biomolecular}
\bibinfo{author}{\bibfnamefont{S.~F.} \bibnamefont{Banani}},
  \bibinfo{author}{\bibfnamefont{H.~O.} \bibnamefont{Lee}},
  \bibinfo{author}{\bibfnamefont{A.~A.} \bibnamefont{Hyman}}, \bibnamefont{and}
  \bibinfo{author}{\bibfnamefont{M.~K.} \bibnamefont{Rosen}},
  \bibinfo{journal}{Nature reviews Molecular cell biology}
  \textbf{\bibinfo{volume}{18}}, \bibinfo{pages}{285} (\bibinfo{year}{2017}).

\bibitem[{\citenamefont{Boeynaems et~al.}(2018)\citenamefont{Boeynaems,
  Alberti, Fawzi, Mittag, Polymenidou, Rousseau, Schymkowitz, Shorter, Wolozin,
  Van Den~Bosch et~al.}}]{boeynaems2018protein}
\bibinfo{author}{\bibfnamefont{S.}~\bibnamefont{Boeynaems}},
  \bibinfo{author}{\bibfnamefont{S.}~\bibnamefont{Alberti}},
  \bibinfo{author}{\bibfnamefont{N.~L.} \bibnamefont{Fawzi}},
  \bibinfo{author}{\bibfnamefont{T.}~\bibnamefont{Mittag}},
  \bibinfo{author}{\bibfnamefont{M.}~\bibnamefont{Polymenidou}},
  \bibinfo{author}{\bibfnamefont{F.}~\bibnamefont{Rousseau}},
  \bibinfo{author}{\bibfnamefont{J.}~\bibnamefont{Schymkowitz}},
  \bibinfo{author}{\bibfnamefont{J.}~\bibnamefont{Shorter}},
  \bibinfo{author}{\bibfnamefont{B.}~\bibnamefont{Wolozin}},
  \bibinfo{author}{\bibfnamefont{L.}~\bibnamefont{Van Den~Bosch}},
  \bibnamefont{et~al.}, \bibinfo{journal}{Trends in cell biology}
  \textbf{\bibinfo{volume}{28}}, \bibinfo{pages}{420} (\bibinfo{year}{2018}).

\bibitem[{\citenamefont{Brangwynne et~al.}(2009)\citenamefont{Brangwynne,
  Eckmann, Courson, Rybarska, Hoege, Gharakhani, J{\"u}licher, and
  Hyman}}]{brangwynne2009germline}
\bibinfo{author}{\bibfnamefont{C.~P.} \bibnamefont{Brangwynne}},
  \bibinfo{author}{\bibfnamefont{C.~R.} \bibnamefont{Eckmann}},
  \bibinfo{author}{\bibfnamefont{D.~S.} \bibnamefont{Courson}},
  \bibinfo{author}{\bibfnamefont{A.}~\bibnamefont{Rybarska}},
  \bibinfo{author}{\bibfnamefont{C.}~\bibnamefont{Hoege}},
  \bibinfo{author}{\bibfnamefont{J.}~\bibnamefont{Gharakhani}},
  \bibinfo{author}{\bibfnamefont{F.}~\bibnamefont{J{\"u}licher}},
  \bibnamefont{and} \bibinfo{author}{\bibfnamefont{A.~A.} \bibnamefont{Hyman}},
  \bibinfo{journal}{Science} \textbf{\bibinfo{volume}{324}},
  \bibinfo{pages}{1729} (\bibinfo{year}{2009}).

\bibitem[{\citenamefont{Kato et~al.}(2012)\citenamefont{Kato, Han, Xie, Shi,
  Du, Wu, Mirzaei, Goldsmith, Longgood, Pei et~al.}}]{Kato2012CellfreeFO}
\bibinfo{author}{\bibfnamefont{M.}~\bibnamefont{Kato}},
  \bibinfo{author}{\bibfnamefont{T.~W.} \bibnamefont{Han}},
  \bibinfo{author}{\bibfnamefont{S.}~\bibnamefont{Xie}},
  \bibinfo{author}{\bibfnamefont{K.}~\bibnamefont{Shi}},
  \bibinfo{author}{\bibfnamefont{X.}~\bibnamefont{Du}},
  \bibinfo{author}{\bibfnamefont{L.~C.} \bibnamefont{Wu}},
  \bibinfo{author}{\bibfnamefont{H.}~\bibnamefont{Mirzaei}},
  \bibinfo{author}{\bibfnamefont{E.~J.} \bibnamefont{Goldsmith}},
  \bibinfo{author}{\bibfnamefont{J.~C.} \bibnamefont{Longgood}},
  \bibinfo{author}{\bibfnamefont{J.}~\bibnamefont{Pei}}, \bibnamefont{et~al.},
  \bibinfo{journal}{Cell} \textbf{\bibinfo{volume}{149}}, \bibinfo{pages}{753}
  (\bibinfo{year}{2012}).

\bibitem[{\citenamefont{Nott et~al.}(2015)\citenamefont{Nott, Petsalaki,
  Farber, Jervis, Fussner, Plochowietz, Craggs, Bazett-Jones, Pawson,
  Forman-Kay et~al.}}]{Nott2015PhaseTO}
\bibinfo{author}{\bibfnamefont{T.~J.} \bibnamefont{Nott}},
  \bibinfo{author}{\bibfnamefont{E.}~\bibnamefont{Petsalaki}},
  \bibinfo{author}{\bibfnamefont{P.~J.} \bibnamefont{Farber}},
  \bibinfo{author}{\bibfnamefont{D.}~\bibnamefont{Jervis}},
  \bibinfo{author}{\bibfnamefont{E.}~\bibnamefont{Fussner}},
  \bibinfo{author}{\bibfnamefont{A.}~\bibnamefont{Plochowietz}},
  \bibinfo{author}{\bibfnamefont{T.~D.} \bibnamefont{Craggs}},
  \bibinfo{author}{\bibfnamefont{D.~P.} \bibnamefont{Bazett-Jones}},
  \bibinfo{author}{\bibfnamefont{T.}~\bibnamefont{Pawson}},
  \bibinfo{author}{\bibfnamefont{J.~D.} \bibnamefont{Forman-Kay}},
  \bibnamefont{et~al.}, \bibinfo{journal}{Molecular Cell}
  \textbf{\bibinfo{volume}{57}}, \bibinfo{pages}{936 } (\bibinfo{year}{2015}).

\bibitem[{\citenamefont{Molliex et~al.}(2015)\citenamefont{Molliex, Temirov,
  Lee, Coughlin, Kanagaraj, Kim, Mittag, and Taylor}}]{molliex2015phase}
\bibinfo{author}{\bibfnamefont{A.}~\bibnamefont{Molliex}},
  \bibinfo{author}{\bibfnamefont{J.}~\bibnamefont{Temirov}},
  \bibinfo{author}{\bibfnamefont{J.}~\bibnamefont{Lee}},
  \bibinfo{author}{\bibfnamefont{M.}~\bibnamefont{Coughlin}},
  \bibinfo{author}{\bibfnamefont{A.~P.} \bibnamefont{Kanagaraj}},
  \bibinfo{author}{\bibfnamefont{H.~J.} \bibnamefont{Kim}},
  \bibinfo{author}{\bibfnamefont{T.}~\bibnamefont{Mittag}}, \bibnamefont{and}
  \bibinfo{author}{\bibfnamefont{J.~P.} \bibnamefont{Taylor}},
  \bibinfo{journal}{Cell} \textbf{\bibinfo{volume}{163}}, \bibinfo{pages}{123}
  (\bibinfo{year}{2015}).

\bibitem[{\citenamefont{Brangwynne et~al.}(2011)\citenamefont{Brangwynne,
  Mitchison, and Hyman}}]{Brangwynne2011ActiveLB}
\bibinfo{author}{\bibfnamefont{C.~P.} \bibnamefont{Brangwynne}},
  \bibinfo{author}{\bibfnamefont{T.~J.} \bibnamefont{Mitchison}},
  \bibnamefont{and} \bibinfo{author}{\bibfnamefont{A.~A.} \bibnamefont{Hyman}},
  \bibinfo{journal}{Proceedings of the National Academy of Sciences}
  \textbf{\bibinfo{volume}{108}}, \bibinfo{pages}{4334 }
  (\bibinfo{year}{2011}).

\bibitem[{\citenamefont{Schuster et~al.}(2021)\citenamefont{Schuster, Regy,
  Dolan, Ranganath, Jovic, Khare, Shi, and Mittal}}]{SchusterLLPSReview2021}
\bibinfo{author}{\bibfnamefont{B.~S.} \bibnamefont{Schuster}},
  \bibinfo{author}{\bibfnamefont{R.~M.} \bibnamefont{Regy}},
  \bibinfo{author}{\bibfnamefont{E.~M.} \bibnamefont{Dolan}},
  \bibinfo{author}{\bibfnamefont{A.~K.} \bibnamefont{Ranganath}},
  \bibinfo{author}{\bibfnamefont{N.}~\bibnamefont{Jovic}},
  \bibinfo{author}{\bibfnamefont{S.~D.} \bibnamefont{Khare}},
  \bibinfo{author}{\bibfnamefont{Z.}~\bibnamefont{Shi}}, \bibnamefont{and}
  \bibinfo{author}{\bibfnamefont{J.}~\bibnamefont{Mittal}},
  \bibinfo{journal}{The journal of physical chemistry. B}
  (\bibinfo{year}{2021}).

\bibitem[{\citenamefont{Lafontaine et~al.}(2020)\citenamefont{Lafontaine,
  Riback, Bascetin, and Brangwynne}}]{LaFontaineNucleolus2020}
\bibinfo{author}{\bibfnamefont{D.~L.~J.} \bibnamefont{Lafontaine}},
  \bibinfo{author}{\bibfnamefont{J.~A.} \bibnamefont{Riback}},
  \bibinfo{author}{\bibfnamefont{R.}~\bibnamefont{Bascetin}}, \bibnamefont{and}
  \bibinfo{author}{\bibfnamefont{C.~P.} \bibnamefont{Brangwynne}},
  \bibinfo{journal}{Nature Reviews Molecular Cell Biology} pp.
  \bibinfo{pages}{1--18} (\bibinfo{year}{2020}).

\bibitem[{\citenamefont{Shin and Brangwynne}(2017)}]{ShinLLPS2017}
\bibinfo{author}{\bibfnamefont{Y.}~\bibnamefont{Shin}} \bibnamefont{and}
  \bibinfo{author}{\bibfnamefont{C.~P.} \bibnamefont{Brangwynne}},
  \bibinfo{journal}{Science} \textbf{\bibinfo{volume}{357}},
  \bibinfo{pages}{eaaf4382} (\bibinfo{year}{2017}),
  \urlprefix\url{https://www.science.org/doi/abs/10.1126/science.aaf4382}.

\bibitem[{\citenamefont{Patel et~al.}(2015)\citenamefont{Patel, Lee, Jawerth,
  Maharana, Jahnel, Hein, Stoynov, Mahamid, Saha, Franzmann
  et~al.}}]{ALSFUS2015}
\bibinfo{author}{\bibfnamefont{A.}~\bibnamefont{Patel}},
  \bibinfo{author}{\bibfnamefont{H.}~\bibnamefont{Lee}},
  \bibinfo{author}{\bibfnamefont{L.}~\bibnamefont{Jawerth}},
  \bibinfo{author}{\bibfnamefont{S.}~\bibnamefont{Maharana}},
  \bibinfo{author}{\bibfnamefont{M.}~\bibnamefont{Jahnel}},
  \bibinfo{author}{\bibfnamefont{M.}~\bibnamefont{Hein}},
  \bibinfo{author}{\bibfnamefont{S.}~\bibnamefont{Stoynov}},
  \bibinfo{author}{\bibfnamefont{J.}~\bibnamefont{Mahamid}},
  \bibinfo{author}{\bibfnamefont{S.}~\bibnamefont{Saha}},
  \bibinfo{author}{\bibfnamefont{T.}~\bibnamefont{Franzmann}},
  \bibnamefont{et~al.}, \bibinfo{journal}{Cell} \textbf{\bibinfo{volume}{162}},
  \bibinfo{pages}{1066} (\bibinfo{year}{2015}), ISSN \bibinfo{issn}{0092-8674},
  \urlprefix\url{https://www.sciencedirect.com/science/article/pii/S0092867415009630}.

\bibitem[{\citenamefont{Bracha et~al.}(2019)\citenamefont{Bracha, Walls, and
  Brangwynne}}]{BrachaProbing2019}
\bibinfo{author}{\bibfnamefont{D.}~\bibnamefont{Bracha}},
  \bibinfo{author}{\bibfnamefont{M.~T.} \bibnamefont{Walls}}, \bibnamefont{and}
  \bibinfo{author}{\bibfnamefont{C.~P.} \bibnamefont{Brangwynne}},
  \bibinfo{journal}{Nature Biotechnology} \textbf{\bibinfo{volume}{37}},
  \bibinfo{pages}{1435} (\bibinfo{year}{2019}),
  \urlprefix\url{https://doi.org/10.1038/s41587-019-0341-6}.

\bibitem[{\citenamefont{Feric et~al.}(2016)\citenamefont{Feric, Vaidya, Harmon,
  Mitrea, Zhu, Richardson, Kriwacki, Pappu, and
  Brangwynne}}]{feric2016coexisting}
\bibinfo{author}{\bibfnamefont{M.}~\bibnamefont{Feric}},
  \bibinfo{author}{\bibfnamefont{N.}~\bibnamefont{Vaidya}},
  \bibinfo{author}{\bibfnamefont{T.~S.} \bibnamefont{Harmon}},
  \bibinfo{author}{\bibfnamefont{D.~M.} \bibnamefont{Mitrea}},
  \bibinfo{author}{\bibfnamefont{L.}~\bibnamefont{Zhu}},
  \bibinfo{author}{\bibfnamefont{T.~M.} \bibnamefont{Richardson}},
  \bibinfo{author}{\bibfnamefont{R.~W.} \bibnamefont{Kriwacki}},
  \bibinfo{author}{\bibfnamefont{R.~V.} \bibnamefont{Pappu}}, \bibnamefont{and}
  \bibinfo{author}{\bibfnamefont{C.~P.} \bibnamefont{Brangwynne}},
  \bibinfo{journal}{Cell} \textbf{\bibinfo{volume}{165}}, \bibinfo{pages}{1686}
  (\bibinfo{year}{2016}).

\bibitem[{\citenamefont{Wang et~al.}(2021)\citenamefont{Wang, Kelley,
  Milovanovic, Schuster, and Shi}}]{WANGaspiration20211}
\bibinfo{author}{\bibfnamefont{H.}~\bibnamefont{Wang}},
  \bibinfo{author}{\bibfnamefont{F.~M.} \bibnamefont{Kelley}},
  \bibinfo{author}{\bibfnamefont{D.}~\bibnamefont{Milovanovic}},
  \bibinfo{author}{\bibfnamefont{B.~S.} \bibnamefont{Schuster}},
  \bibnamefont{and} \bibinfo{author}{\bibfnamefont{Z.}~\bibnamefont{Shi}},
  \bibinfo{journal}{Biophysical Reports} \textbf{\bibinfo{volume}{1}}
  (\bibinfo{year}{2021}).

\bibitem[{\citenamefont{Ijavi et~al.}(2020)\citenamefont{Ijavi, Style,
  Emmanouilidis, Kumar, Meier, Torzynski, Allain, Barral, Steinmetz, and
  Dufresne}}]{ijavi2021surface}
\bibinfo{author}{\bibfnamefont{M.}~\bibnamefont{Ijavi}},
  \bibinfo{author}{\bibfnamefont{R.~W.} \bibnamefont{Style}},
  \bibinfo{author}{\bibfnamefont{L.}~\bibnamefont{Emmanouilidis}},
  \bibinfo{author}{\bibfnamefont{A.}~\bibnamefont{Kumar}},
  \bibinfo{author}{\bibfnamefont{S.~M.} \bibnamefont{Meier}},
  \bibinfo{author}{\bibfnamefont{A.~L.} \bibnamefont{Torzynski}},
  \bibinfo{author}{\bibfnamefont{F.~H.-T.} \bibnamefont{Allain}},
  \bibinfo{author}{\bibfnamefont{Y.}~\bibnamefont{Barral}},
  \bibinfo{author}{\bibfnamefont{M.~O.} \bibnamefont{Steinmetz}},
  \bibnamefont{and} \bibinfo{author}{\bibfnamefont{E.~R.}
  \bibnamefont{Dufresne}}, \bibinfo{journal}{Soft matter}
  (\bibinfo{year}{2020}).

\bibitem[{\citenamefont{Alshareedah et~al.}(2021)\citenamefont{Alshareedah,
  Thurston, and Banerjee}}]{FusionOT2021}
\bibinfo{author}{\bibfnamefont{I.}~\bibnamefont{Alshareedah}},
  \bibinfo{author}{\bibfnamefont{G.~M.} \bibnamefont{Thurston}},
  \bibnamefont{and} \bibinfo{author}{\bibfnamefont{P.~R.}
  \bibnamefont{Banerjee}}, \bibinfo{journal}{Biophysical journal}
  (\bibinfo{year}{2021}).

\bibitem[{\citenamefont{Jawerth et~al.}(2020)\citenamefont{Jawerth,
  Fischer-Friedrich, Saha, Wang, Franzmann, Zhang, Sachweh, Ruer, Ijavi, Saha
  et~al.}}]{JawerthOT}
\bibinfo{author}{\bibfnamefont{L.}~\bibnamefont{Jawerth}},
  \bibinfo{author}{\bibfnamefont{E.}~\bibnamefont{Fischer-Friedrich}},
  \bibinfo{author}{\bibfnamefont{S.}~\bibnamefont{Saha}},
  \bibinfo{author}{\bibfnamefont{J.}~\bibnamefont{Wang}},
  \bibinfo{author}{\bibfnamefont{T.}~\bibnamefont{Franzmann}},
  \bibinfo{author}{\bibfnamefont{X.}~\bibnamefont{Zhang}},
  \bibinfo{author}{\bibfnamefont{J.}~\bibnamefont{Sachweh}},
  \bibinfo{author}{\bibfnamefont{M.}~\bibnamefont{Ruer}},
  \bibinfo{author}{\bibfnamefont{M.}~\bibnamefont{Ijavi}},
  \bibinfo{author}{\bibfnamefont{S.}~\bibnamefont{Saha}}, \bibnamefont{et~al.},
  \bibinfo{journal}{Science} \textbf{\bibinfo{volume}{370}},
  \bibinfo{pages}{1317} (\bibinfo{year}{2020}),
  \urlprefix\url{https://www.science.org/doi/abs/10.1126/science.aaw4951}.

\bibitem[{\citenamefont{Shafrin and Zisman}(1960)}]{WettingSurfaceTension1960}
\bibinfo{author}{\bibfnamefont{E.~G.} \bibnamefont{Shafrin}} \bibnamefont{and}
  \bibinfo{author}{\bibfnamefont{W.~A.} \bibnamefont{Zisman}},
  \bibinfo{journal}{The Journal of Physical Chemistry}
  \textbf{\bibinfo{volume}{64}}, \bibinfo{pages}{519} (\bibinfo{year}{1960}).

\bibitem[{\citenamefont{Eggers et~al.}(1999)\citenamefont{Eggers, Lister, and
  Stone}}]{eggersDrops1999}
\bibinfo{author}{\bibfnamefont{J.~G.} \bibnamefont{Eggers}},
  \bibinfo{author}{\bibfnamefont{J.~R.} \bibnamefont{Lister}},
  \bibnamefont{and} \bibinfo{author}{\bibfnamefont{H.~A.} \bibnamefont{Stone}},
  \bibinfo{journal}{Journal of Fluid Mechanics} \textbf{\bibinfo{volume}{401}},
  \bibinfo{pages}{293 } (\bibinfo{year}{1999}).

\bibitem[{\citenamefont{Ellis}(2001)}]{ellis2001macromolecular}
\bibinfo{author}{\bibfnamefont{R.~J.} \bibnamefont{Ellis}},
  \bibinfo{journal}{Trends in biochemical sciences}
  \textbf{\bibinfo{volume}{26}}, \bibinfo{pages}{597} (\bibinfo{year}{2001}).

\bibitem[{\citenamefont{Hoppe and Minton}(2019)}]{hoppe2019non}
\bibinfo{author}{\bibfnamefont{T.}~\bibnamefont{Hoppe}} \bibnamefont{and}
  \bibinfo{author}{\bibfnamefont{A.~P.} \bibnamefont{Minton}},
  \bibinfo{journal}{Frontiers in molecular biosciences}
  \textbf{\bibinfo{volume}{6}}, \bibinfo{pages}{10} (\bibinfo{year}{2019}).

\bibitem[{\citenamefont{Chen et~al.}(2022)\citenamefont{Chen, Wang, Tian,
  Zhang, Song, Crick, Carmalt, Parkin, and Lu}}]{Repellent2022}
\bibinfo{author}{\bibfnamefont{F.}~\bibnamefont{Chen}},
  \bibinfo{author}{\bibfnamefont{Y.}~\bibnamefont{Wang}},
  \bibinfo{author}{\bibfnamefont{Y.}~\bibnamefont{Tian}},
  \bibinfo{author}{\bibfnamefont{D.}~\bibnamefont{Zhang}},
  \bibinfo{author}{\bibfnamefont{J.}~\bibnamefont{Song}},
  \bibinfo{author}{\bibfnamefont{C.~R.} \bibnamefont{Crick}},
  \bibinfo{author}{\bibfnamefont{C.~J.} \bibnamefont{Carmalt}},
  \bibinfo{author}{\bibfnamefont{I.~P.} \bibnamefont{Parkin}},
  \bibnamefont{and} \bibinfo{author}{\bibfnamefont{Y.}~\bibnamefont{Lu}},
  \bibinfo{journal}{Chem. Soc. Rev.} \textbf{\bibinfo{volume}{51}},
  \bibinfo{pages}{8476} (\bibinfo{year}{2022}),
  \urlprefix\url{http://dx.doi.org/10.1039/D0CS01033B}.

\bibitem[{\citenamefont{Chen et~al.}(2020)\citenamefont{Chen, Cui, Han, Hu,
  Sun, Zhang, Wang, Song, Chen, and Lou}}]{chen2020liquid}
\bibinfo{author}{\bibfnamefont{H.}~\bibnamefont{Chen}},
  \bibinfo{author}{\bibfnamefont{Y.}~\bibnamefont{Cui}},
  \bibinfo{author}{\bibfnamefont{X.}~\bibnamefont{Han}},
  \bibinfo{author}{\bibfnamefont{W.}~\bibnamefont{Hu}},
  \bibinfo{author}{\bibfnamefont{M.}~\bibnamefont{Sun}},
  \bibinfo{author}{\bibfnamefont{Y.}~\bibnamefont{Zhang}},
  \bibinfo{author}{\bibfnamefont{P.-H.} \bibnamefont{Wang}},
  \bibinfo{author}{\bibfnamefont{G.}~\bibnamefont{Song}},
  \bibinfo{author}{\bibfnamefont{W.}~\bibnamefont{Chen}}, \bibnamefont{and}
  \bibinfo{author}{\bibfnamefont{J.}~\bibnamefont{Lou}}, \bibinfo{journal}{Cell
  research} \textbf{\bibinfo{volume}{30}}, \bibinfo{pages}{1143}
  (\bibinfo{year}{2020}).

\bibitem[{\citenamefont{Lin et~al.}(2015)\citenamefont{Lin, Protter, Rosen, and
  Parker}}]{lin2015formation}
\bibinfo{author}{\bibfnamefont{Y.}~\bibnamefont{Lin}},
  \bibinfo{author}{\bibfnamefont{D.~S.} \bibnamefont{Protter}},
  \bibinfo{author}{\bibfnamefont{M.~K.} \bibnamefont{Rosen}}, \bibnamefont{and}
  \bibinfo{author}{\bibfnamefont{R.}~\bibnamefont{Parker}},
  \bibinfo{journal}{Molecular cell} \textbf{\bibinfo{volume}{60}},
  \bibinfo{pages}{208} (\bibinfo{year}{2015}).

\bibitem[{\citenamefont{Jeon et~al.}(2018)\citenamefont{Jeon, Nguyen, Abraham,
  Conrad, Fygenson, and Saleh}}]{jeon2018salt}
\bibinfo{author}{\bibfnamefont{B.-j.} \bibnamefont{Jeon}},
  \bibinfo{author}{\bibfnamefont{D.~T.} \bibnamefont{Nguyen}},
  \bibinfo{author}{\bibfnamefont{G.~R.} \bibnamefont{Abraham}},
  \bibinfo{author}{\bibfnamefont{N.}~\bibnamefont{Conrad}},
  \bibinfo{author}{\bibfnamefont{D.~K.} \bibnamefont{Fygenson}},
  \bibnamefont{and} \bibinfo{author}{\bibfnamefont{O.~A.} \bibnamefont{Saleh}},
  \bibinfo{journal}{Soft Matter} \textbf{\bibinfo{volume}{14}},
  \bibinfo{pages}{7009} (\bibinfo{year}{2018}).

\bibitem[{\citenamefont{Wang and McCarthy}(2016)}]{SOCAL2016}
\bibinfo{author}{\bibfnamefont{L.}~\bibnamefont{Wang}} \bibnamefont{and}
  \bibinfo{author}{\bibfnamefont{T.~J.} \bibnamefont{McCarthy}},
  \bibinfo{journal}{Angewandte Chemie International Edition}
  \textbf{\bibinfo{volume}{55}}, \bibinfo{pages}{244} (\bibinfo{year}{2016}),
  \urlprefix\url{https://onlinelibrary.wiley.com/doi/abs/10.1002/anie.201509385}.

\bibitem[{\citenamefont{Sofia et~al.}(1998)\citenamefont{Sofia, Premnath, and
  Merrill}}]{Sofia1998PolyethyleneOG}
\bibinfo{author}{\bibnamefont{Sofia}},
  \bibinfo{author}{\bibnamefont{Premnath}}, \bibnamefont{and}
  \bibinfo{author}{\bibnamefont{Merrill}}, \bibinfo{journal}{Macromolecules}
  \textbf{\bibinfo{volume}{31 15}}, \bibinfo{pages}{5059}
  (\bibinfo{year}{1998}).

\bibitem[{\citenamefont{Upadhyayula et~al.}(2012)\citenamefont{Upadhyayula,
  Quinata, Bishop, Gupta, Johnson, Bahmani, Bozhilov, Stubbs, Jreij, Nallagatla
  et~al.}}]{Upadhyayula2012CoatingsOP}
\bibinfo{author}{\bibfnamefont{S.}~\bibnamefont{Upadhyayula}},
  \bibinfo{author}{\bibfnamefont{T.}~\bibnamefont{Quinata}},
  \bibinfo{author}{\bibfnamefont{S.}~\bibnamefont{Bishop}},
  \bibinfo{author}{\bibfnamefont{S.}~\bibnamefont{Gupta}},
  \bibinfo{author}{\bibfnamefont{N.~R.} \bibnamefont{Johnson}},
  \bibinfo{author}{\bibfnamefont{B.}~\bibnamefont{Bahmani}},
  \bibinfo{author}{\bibfnamefont{K.}~\bibnamefont{Bozhilov}},
  \bibinfo{author}{\bibfnamefont{J.}~\bibnamefont{Stubbs}},
  \bibinfo{author}{\bibfnamefont{P.}~\bibnamefont{Jreij}},
  \bibinfo{author}{\bibfnamefont{P.}~\bibnamefont{Nallagatla}},
  \bibnamefont{et~al.}, \bibinfo{journal}{Langmuir : the ACS journal of
  surfaces and colloids} \textbf{\bibinfo{volume}{28 11}},
  \bibinfo{pages}{5059} (\bibinfo{year}{2012}).

\bibitem[{\citenamefont{Kirkness et~al.}(2018)\citenamefont{Kirkness, Korosec,
  and Forde}}]{kirkness2018modified}
\bibinfo{author}{\bibfnamefont{M.~W.} \bibnamefont{Kirkness}},
  \bibinfo{author}{\bibfnamefont{C.~S.} \bibnamefont{Korosec}},
  \bibnamefont{and} \bibinfo{author}{\bibfnamefont{N.~R.} \bibnamefont{Forde}},
  \bibinfo{journal}{Langmuir} \textbf{\bibinfo{volume}{34}},
  \bibinfo{pages}{13550} (\bibinfo{year}{2018}).

\bibitem[{\citenamefont{Wong et~al.}(2011{\natexlab{a}})\citenamefont{Wong,
  Kang, Tang, Smythe, Hatton, Grinthal, and Aizenberg}}]{SLIPSnature2011}
\bibinfo{author}{\bibfnamefont{T.-S.} \bibnamefont{Wong}},
  \bibinfo{author}{\bibfnamefont{S.~H.} \bibnamefont{Kang}},
  \bibinfo{author}{\bibfnamefont{S.~K.~Y.} \bibnamefont{Tang}},
  \bibinfo{author}{\bibfnamefont{E.~J.} \bibnamefont{Smythe}},
  \bibinfo{author}{\bibfnamefont{B.~D.} \bibnamefont{Hatton}},
  \bibinfo{author}{\bibfnamefont{A.}~\bibnamefont{Grinthal}}, \bibnamefont{and}
  \bibinfo{author}{\bibfnamefont{J.}~\bibnamefont{Aizenberg}},
  \bibinfo{journal}{Nature} \textbf{\bibinfo{volume}{477}}, \bibinfo{pages}{443
  } (\bibinfo{year}{2011}{\natexlab{a}}),
  \urlprefix\url{https://doi.org/10.1038/nature10447}.

\bibitem[{\citenamefont{Spanke et~al.}(2020)\citenamefont{Spanke, Style,
  Fran{\c{c}}ois-Martin, Feofilova, Eisentraut, Kress, Agudo-Canalejo, and
  Dufresne}}]{spanke2020wrapping}
\bibinfo{author}{\bibfnamefont{H.~T.} \bibnamefont{Spanke}},
  \bibinfo{author}{\bibfnamefont{R.~W.} \bibnamefont{Style}},
  \bibinfo{author}{\bibfnamefont{C.}~\bibnamefont{Fran{\c{c}}ois-Martin}},
  \bibinfo{author}{\bibfnamefont{M.}~\bibnamefont{Feofilova}},
  \bibinfo{author}{\bibfnamefont{M.}~\bibnamefont{Eisentraut}},
  \bibinfo{author}{\bibfnamefont{H.}~\bibnamefont{Kress}},
  \bibinfo{author}{\bibfnamefont{J.}~\bibnamefont{Agudo-Canalejo}},
  \bibnamefont{and} \bibinfo{author}{\bibfnamefont{E.~R.}
  \bibnamefont{Dufresne}}, \bibinfo{journal}{Physical Review Letters}
  \textbf{\bibinfo{volume}{125}}, \bibinfo{pages}{198102}
  (\bibinfo{year}{2020}).

\bibitem[{\citenamefont{Testa et~al.}(2021)\citenamefont{Testa, Dindo, Rebane,
  Nasouri, Style, Golestanian, Dufresne, and Laurino}}]{testa2021sustained}
\bibinfo{author}{\bibfnamefont{A.}~\bibnamefont{Testa}},
  \bibinfo{author}{\bibfnamefont{M.}~\bibnamefont{Dindo}},
  \bibinfo{author}{\bibfnamefont{A.~A.} \bibnamefont{Rebane}},
  \bibinfo{author}{\bibfnamefont{B.}~\bibnamefont{Nasouri}},
  \bibinfo{author}{\bibfnamefont{R.~W.} \bibnamefont{Style}},
  \bibinfo{author}{\bibfnamefont{R.}~\bibnamefont{Golestanian}},
  \bibinfo{author}{\bibfnamefont{E.~R.} \bibnamefont{Dufresne}},
  \bibnamefont{and} \bibinfo{author}{\bibfnamefont{P.}~\bibnamefont{Laurino}},
  \bibinfo{journal}{Nature Communications} \textbf{\bibinfo{volume}{12}}
  (\bibinfo{year}{2021}).

\bibitem[{\citenamefont{Bradley et~al.}(2019)\citenamefont{Bradley, Box,
  Hewitt, and Vella}}]{Bradley2019PRL}
\bibinfo{author}{\bibfnamefont{A.~T.} \bibnamefont{Bradley}},
  \bibinfo{author}{\bibfnamefont{F.}~\bibnamefont{Box}},
  \bibinfo{author}{\bibfnamefont{I.~J.} \bibnamefont{Hewitt}},
  \bibnamefont{and} \bibinfo{author}{\bibfnamefont{D.}~\bibnamefont{Vella}},
  \bibinfo{journal}{Phys. Rev. Lett.} \textbf{\bibinfo{volume}{122}},
  \bibinfo{pages}{074503} (\bibinfo{year}{2019}),
  \urlprefix\url{https://link.aps.org/doi/10.1103/PhysRevLett.122.074503}.

\bibitem[{\citenamefont{Drelich et~al.}(2020)\citenamefont{Drelich, Boinovich,
  Chibowski, Della~Volpe, Ho\l{}ysz, Marmur, and
  Siboni}}]{Drelich2020contactangles}
\bibinfo{author}{\bibfnamefont{J.~W.} \bibnamefont{Drelich}},
  \bibinfo{author}{\bibfnamefont{L.}~\bibnamefont{Boinovich}},
  \bibinfo{author}{\bibfnamefont{E.}~\bibnamefont{Chibowski}},
  \bibinfo{author}{\bibfnamefont{C.}~\bibnamefont{Della~Volpe}},
  \bibinfo{author}{\bibfnamefont{L.}~\bibnamefont{Ho\l{}ysz}},
  \bibinfo{author}{\bibfnamefont{A.}~\bibnamefont{Marmur}}, \bibnamefont{and}
  \bibinfo{author}{\bibfnamefont{S.}~\bibnamefont{Siboni}},
  \bibinfo{journal}{Surface Innovations} \textbf{\bibinfo{volume}{8}},
  \bibinfo{pages}{3} (\bibinfo{year}{2020}),
  \urlprefix\url{https://doi.org/10.1680/jsuin.19.00007}.

\bibitem[{\citenamefont{Chu and Seeger}(2014)}]{Chu2014CSR}
\bibinfo{author}{\bibfnamefont{Z.}~\bibnamefont{Chu}} \bibnamefont{and}
  \bibinfo{author}{\bibfnamefont{S.}~\bibnamefont{Seeger}},
  \bibinfo{journal}{Chem. Soc. Rev.} \textbf{\bibinfo{volume}{43}},
  \bibinfo{pages}{2784} (\bibinfo{year}{2014}),
  \urlprefix\url{http://dx.doi.org/10.1039/C3CS60415B}.

\bibitem[{\citenamefont{Faltova et~al.}(2018)\citenamefont{Faltova,
  K\"{u}ffner, Hondele, Weis, and Arosio}}]{faltova2018multifunctional}
\bibinfo{author}{\bibfnamefont{L.}~\bibnamefont{Faltova}},
  \bibinfo{author}{\bibfnamefont{A.~M.} \bibnamefont{K\"{u}ffner}},
  \bibinfo{author}{\bibfnamefont{M.}~\bibnamefont{Hondele}},
  \bibinfo{author}{\bibfnamefont{K.}~\bibnamefont{Weis}}, \bibnamefont{and}
  \bibinfo{author}{\bibfnamefont{P.}~\bibnamefont{Arosio}},
  \bibinfo{journal}{ACS nano} \textbf{\bibinfo{volume}{12}},
  \bibinfo{pages}{9991} (\bibinfo{year}{2018}).

\bibitem[{\citenamefont{K{\"u}ffner et~al.}(2020)\citenamefont{K{\"u}ffner,
  Prodan, Zuccarini, Capasso~Palmiero, Faltova, and
  Arosio}}]{kuffner2020acceleration}
\bibinfo{author}{\bibfnamefont{A.~M.} \bibnamefont{K{\"u}ffner}},
  \bibinfo{author}{\bibfnamefont{M.}~\bibnamefont{Prodan}},
  \bibinfo{author}{\bibfnamefont{R.}~\bibnamefont{Zuccarini}},
  \bibinfo{author}{\bibfnamefont{U.}~\bibnamefont{Capasso~Palmiero}},
  \bibinfo{author}{\bibfnamefont{L.}~\bibnamefont{Faltova}}, \bibnamefont{and}
  \bibinfo{author}{\bibfnamefont{P.}~\bibnamefont{Arosio}},
  \bibinfo{journal}{ChemSystemsChem} \textbf{\bibinfo{volume}{2}},
  \bibinfo{pages}{e2000001} (\bibinfo{year}{2020}).

\bibitem[{\citenamefont{Klepinin et~al.}(2020)\citenamefont{Klepinin, Zhang,
  Klepinina, Rebane-Klemm, Terzic, Kaambre, and Dzeja}}]{klepinin2020adenylate}
\bibinfo{author}{\bibfnamefont{A.}~\bibnamefont{Klepinin}},
  \bibinfo{author}{\bibfnamefont{S.}~\bibnamefont{Zhang}},
  \bibinfo{author}{\bibfnamefont{L.}~\bibnamefont{Klepinina}},
  \bibinfo{author}{\bibfnamefont{E.}~\bibnamefont{Rebane-Klemm}},
  \bibinfo{author}{\bibfnamefont{A.}~\bibnamefont{Terzic}},
  \bibinfo{author}{\bibfnamefont{T.}~\bibnamefont{Kaambre}}, \bibnamefont{and}
  \bibinfo{author}{\bibfnamefont{P.}~\bibnamefont{Dzeja}},
  \bibinfo{journal}{Frontiers in oncology} \textbf{\bibinfo{volume}{10}},
  \bibinfo{pages}{660} (\bibinfo{year}{2020}).

\bibitem[{\citenamefont{Elbaum-Garfinkle
  et~al.}(2015)\citenamefont{Elbaum-Garfinkle, Kim, Szczepaniak, Chen, Eckmann,
  Myong, and Brangwynne}}]{elbaum2015disordered}
\bibinfo{author}{\bibfnamefont{S.}~\bibnamefont{Elbaum-Garfinkle}},
  \bibinfo{author}{\bibfnamefont{Y.}~\bibnamefont{Kim}},
  \bibinfo{author}{\bibfnamefont{K.}~\bibnamefont{Szczepaniak}},
  \bibinfo{author}{\bibfnamefont{C.~C.-H.} \bibnamefont{Chen}},
  \bibinfo{author}{\bibfnamefont{C.~R.} \bibnamefont{Eckmann}},
  \bibinfo{author}{\bibfnamefont{S.}~\bibnamefont{Myong}}, \bibnamefont{and}
  \bibinfo{author}{\bibfnamefont{C.~P.} \bibnamefont{Brangwynne}},
  \bibinfo{journal}{Proceedings of the National Academy of Sciences}
  \textbf{\bibinfo{volume}{112}}, \bibinfo{pages}{7189} (\bibinfo{year}{2015}).

\bibitem[{\citenamefont{Brangwynne et~al.}(2015)\citenamefont{Brangwynne,
  Tompa, and Pappu}}]{Brangwynne2015NatPhys}
\bibinfo{author}{\bibfnamefont{C.~P.} \bibnamefont{Brangwynne}},
  \bibinfo{author}{\bibfnamefont{P.}~\bibnamefont{Tompa}}, \bibnamefont{and}
  \bibinfo{author}{\bibfnamefont{R.~V.} \bibnamefont{Pappu}},
  \bibinfo{journal}{Nature Physics} \textbf{\bibinfo{volume}{11}},
  \bibinfo{pages}{899} (\bibinfo{year}{2015}),
  \urlprefix\url{https://doi.org/10.1038/nphys3532}.

\bibitem[{\citenamefont{Williams et~al.}(2012)\citenamefont{Williams, Koga,
  Hak, Majrekar, Patil, Perriman, and Mann}}]{williams2012polymer}
\bibinfo{author}{\bibfnamefont{D.~S.} \bibnamefont{Williams}},
  \bibinfo{author}{\bibfnamefont{S.}~\bibnamefont{Koga}},
  \bibinfo{author}{\bibfnamefont{C.~R.~C.} \bibnamefont{Hak}},
  \bibinfo{author}{\bibfnamefont{A.}~\bibnamefont{Majrekar}},
  \bibinfo{author}{\bibfnamefont{A.~J.} \bibnamefont{Patil}},
  \bibinfo{author}{\bibfnamefont{A.~W.} \bibnamefont{Perriman}},
  \bibnamefont{and} \bibinfo{author}{\bibfnamefont{S.}~\bibnamefont{Mann}},
  \bibinfo{journal}{Soft Matter} \textbf{\bibinfo{volume}{8}},
  \bibinfo{pages}{6004} (\bibinfo{year}{2012}).

\bibitem[{\citenamefont{Rumyantsev et~al.}(2021)\citenamefont{Rumyantsev,
  Jackson, and de~Pablo}}]{Rumyantsev2021Annurev}
\bibinfo{author}{\bibfnamefont{A.~M.} \bibnamefont{Rumyantsev}},
  \bibinfo{author}{\bibfnamefont{N.~E.} \bibnamefont{Jackson}},
  \bibnamefont{and} \bibinfo{author}{\bibfnamefont{J.~J.}
  \bibnamefont{de~Pablo}}, \bibinfo{journal}{Annual Review of Condensed Matter
  Physics} \textbf{\bibinfo{volume}{12}}, \bibinfo{pages}{155}
  (\bibinfo{year}{2021}),
  \urlprefix\url{https://doi.org/10.1146/annurev-conmatphys-042020-113457}.

\bibitem[{\citenamefont{Biffi et~al.}(2013)\citenamefont{Biffi, Cerbino,
  Bomboi, Paraboschi, Asselta, Sciortino, and Bellini}}]{Biffi2013PNAS}
\bibinfo{author}{\bibfnamefont{S.}~\bibnamefont{Biffi}},
  \bibinfo{author}{\bibfnamefont{R.}~\bibnamefont{Cerbino}},
  \bibinfo{author}{\bibfnamefont{F.}~\bibnamefont{Bomboi}},
  \bibinfo{author}{\bibfnamefont{E.~M.} \bibnamefont{Paraboschi}},
  \bibinfo{author}{\bibfnamefont{R.}~\bibnamefont{Asselta}},
  \bibinfo{author}{\bibfnamefont{F.}~\bibnamefont{Sciortino}},
  \bibnamefont{and} \bibinfo{author}{\bibfnamefont{T.}~\bibnamefont{Bellini}},
  \bibinfo{journal}{Proc. Natl. Acad. Sci. U. S. A.}
  \textbf{\bibinfo{volume}{110}}, \bibinfo{pages}{15633}
  (\bibinfo{year}{2013}).

\bibitem[{\citenamefont{Spanke et~al.}(2022)\citenamefont{Spanke,
  Agudo-Canalejo, Tran, Style, and Dufresne}}]{spanke2022dynamics}
\bibinfo{author}{\bibfnamefont{H.~T.} \bibnamefont{Spanke}},
  \bibinfo{author}{\bibfnamefont{J.}~\bibnamefont{Agudo-Canalejo}},
  \bibinfo{author}{\bibfnamefont{D.}~\bibnamefont{Tran}},
  \bibinfo{author}{\bibfnamefont{R.~W.} \bibnamefont{Style}}, \bibnamefont{and}
  \bibinfo{author}{\bibfnamefont{E.~R.} \bibnamefont{Dufresne}},
  \bibinfo{journal}{Physical Review Research} \textbf{\bibinfo{volume}{4}},
  \bibinfo{pages}{023080} (\bibinfo{year}{2022}).

\bibitem[{\citenamefont{Walde et~al.}(2010)\citenamefont{Walde, Cosentino,
  Engel, and Stano}}]{walde2010giant}
\bibinfo{author}{\bibfnamefont{P.}~\bibnamefont{Walde}},
  \bibinfo{author}{\bibfnamefont{K.}~\bibnamefont{Cosentino}},
  \bibinfo{author}{\bibfnamefont{H.}~\bibnamefont{Engel}}, \bibnamefont{and}
  \bibinfo{author}{\bibfnamefont{P.}~\bibnamefont{Stano}},
  \bibinfo{journal}{ChemBioChem} \textbf{\bibinfo{volume}{11}},
  \bibinfo{pages}{848} (\bibinfo{year}{2010}).

\bibitem[{\citenamefont{Spruijt et~al.}(2010)\citenamefont{Spruijt, Sprakel,
  Cohen~Stuart, and van~der Gucht}}]{Spruijt2010Soft}
\bibinfo{author}{\bibfnamefont{E.}~\bibnamefont{Spruijt}},
  \bibinfo{author}{\bibfnamefont{J.}~\bibnamefont{Sprakel}},
  \bibinfo{author}{\bibfnamefont{M.~A.} \bibnamefont{Cohen~Stuart}},
  \bibnamefont{and} \bibinfo{author}{\bibfnamefont{J.}~\bibnamefont{van~der
  Gucht}}, \bibinfo{journal}{Soft Matter} \textbf{\bibinfo{volume}{6}},
  \bibinfo{pages}{172} (\bibinfo{year}{2010}),
  \urlprefix\url{http://dx.doi.org/10.1039/B911541B}.

\bibitem[{\citenamefont{Lekkerkerker and
  Tuinier}(2011)}]{Lekkerkerker2011ColloidsAT}
\bibinfo{author}{\bibfnamefont{H.~N.~W.} \bibnamefont{Lekkerkerker}}
  \bibnamefont{and} \bibinfo{author}{\bibfnamefont{R.}~\bibnamefont{Tuinier}}
  (\bibinfo{year}{2011}).

\bibitem[{\citenamefont{Kuznetsova et~al.}(2014)\citenamefont{Kuznetsova,
  Turoverov, and Uversky}}]{UverskyCrowding2014}
\bibinfo{author}{\bibfnamefont{I.~M.} \bibnamefont{Kuznetsova}},
  \bibinfo{author}{\bibfnamefont{K.~K.} \bibnamefont{Turoverov}},
  \bibnamefont{and} \bibinfo{author}{\bibfnamefont{V.~N.}
  \bibnamefont{Uversky}}, \bibinfo{journal}{International Journal of Molecular
  Sciences} \textbf{\bibinfo{volume}{15}}, \bibinfo{pages}{23090}
  (\bibinfo{year}{2014}),
  \urlprefix\url{https://www.mdpi.com/1422-0067/15/12/23090}.

\bibitem[{\citenamefont{Wang and Annunziata}(2007)}]{WangPEG2007}
\bibinfo{author}{\bibfnamefont{Y.}~\bibnamefont{Wang}} \bibnamefont{and}
  \bibinfo{author}{\bibfnamefont{O.}~\bibnamefont{Annunziata}},
  \bibinfo{journal}{The Journal of Physical Chemistry B}
  \textbf{\bibinfo{volume}{111}}, \bibinfo{pages}{1222} (\bibinfo{year}{2007}),
  \bibinfo{note}{pMID: 17266278},
  \urlprefix\url{https://doi.org/10.1021/jp065608u}.

\bibitem[{\citenamefont{Park et~al.}(2020)\citenamefont{Park, Park, Barnes,
  Lin, Jeon, Najafi, , Delaney, Fredrickson, Shea et~al.}}]{HanPEG2020}
\bibinfo{author}{\bibfnamefont{S.}~\bibnamefont{Park}},
  \bibinfo{author}{\bibfnamefont{S.}~\bibnamefont{Park}},
  \bibinfo{author}{\bibfnamefont{R.}~\bibnamefont{Barnes}},
  \bibinfo{author}{\bibfnamefont{Y.}~\bibnamefont{Lin}},
  \bibinfo{author}{\bibfnamefont{B.-j.} \bibnamefont{Jeon}},
  \bibinfo{author}{\bibfnamefont{S.}~\bibnamefont{Najafi}}, ,
  \bibinfo{author}{\bibfnamefont{K.~T.} \bibnamefont{Delaney}},
  \bibinfo{author}{\bibfnamefont{G.~H.} \bibnamefont{Fredrickson}},
  \bibinfo{author}{\bibfnamefont{J.-E.} \bibnamefont{Shea}},
  \bibnamefont{et~al.}, \bibinfo{journal}{Communications Chemistry}
  \textbf{\bibinfo{volume}{3}}, \bibinfo{pages}{83} (\bibinfo{year}{2020}),
  \urlprefix\url{https://doi.org/10.1038/s42004-020-0328-8}.

\bibitem[{\citenamefont{Butt et~al.}(2018)\citenamefont{Butt, Berger, Steffen,
  Vollmer, and Weber}}]{Butt2018AdaptiveWI}
\bibinfo{author}{\bibfnamefont{H.-J.} \bibnamefont{Butt}},
  \bibinfo{author}{\bibfnamefont{R.}~\bibnamefont{Berger}},
  \bibinfo{author}{\bibfnamefont{W.}~\bibnamefont{Steffen}},
  \bibinfo{author}{\bibfnamefont{D.}~\bibnamefont{Vollmer}}, \bibnamefont{and}
  \bibinfo{author}{\bibfnamefont{S.~A.~L.} \bibnamefont{Weber}},
  \bibinfo{journal}{Langmuir : the ACS journal of surfaces and colloids}
  \textbf{\bibinfo{volume}{34 38}}, \bibinfo{pages}{11292}
  (\bibinfo{year}{2018}).

\bibitem[{\citenamefont{Kim et~al.}(2021)\citenamefont{Kim, Heyden, Gerber,
  Bain, Dufresne, and Style}}]{Kim2021PRX}
\bibinfo{author}{\bibfnamefont{J.~Y.} \bibnamefont{Kim}},
  \bibinfo{author}{\bibfnamefont{S.}~\bibnamefont{Heyden}},
  \bibinfo{author}{\bibfnamefont{D.}~\bibnamefont{Gerber}},
  \bibinfo{author}{\bibfnamefont{N.}~\bibnamefont{Bain}},
  \bibinfo{author}{\bibfnamefont{E.~R.} \bibnamefont{Dufresne}},
  \bibnamefont{and} \bibinfo{author}{\bibfnamefont{R.~W.} \bibnamefont{Style}},
  \bibinfo{journal}{Phys. Rev. X} \textbf{\bibinfo{volume}{11}},
  \bibinfo{pages}{031004} (\bibinfo{year}{2021}),
  \urlprefix\url{https://link.aps.org/doi/10.1103/PhysRevX.11.031004}.

\bibitem[{\citenamefont{Shi et~al.}(2018)\citenamefont{Shi, Zhang, Liu, Hanaor,
  and Gan}}]{shi2018dynamic}
\bibinfo{author}{\bibfnamefont{Z.}~\bibnamefont{Shi}},
  \bibinfo{author}{\bibfnamefont{Y.}~\bibnamefont{Zhang}},
  \bibinfo{author}{\bibfnamefont{M.}~\bibnamefont{Liu}},
  \bibinfo{author}{\bibfnamefont{D.~A.} \bibnamefont{Hanaor}},
  \bibnamefont{and} \bibinfo{author}{\bibfnamefont{Y.}~\bibnamefont{Gan}},
  \bibinfo{journal}{Colloids and Surfaces A: Physicochemical and Engineering
  Aspects} \textbf{\bibinfo{volume}{555}}, \bibinfo{pages}{365}
  (\bibinfo{year}{2018}).

\bibitem[{\citenamefont{Eral et~al.}(2013)\citenamefont{Eral, t~Mannetje, and
  Oh}}]{eral2013contact}
\bibinfo{author}{\bibfnamefont{H.}~\bibnamefont{Eral}},
  \bibinfo{author}{\bibfnamefont{D.}~\bibnamefont{t~Mannetje}},
  \bibnamefont{and} \bibinfo{author}{\bibfnamefont{J.~M.} \bibnamefont{Oh}},
  \bibinfo{journal}{Colloid and polymer science}
  \textbf{\bibinfo{volume}{291}}, \bibinfo{pages}{247} (\bibinfo{year}{2013}).

\bibitem[{\citenamefont{Wong et~al.}(2011{\natexlab{b}})\citenamefont{Wong,
  Kang, Tang, Smythe, Hatton, Grinthal, and Aizenberg}}]{wong2011bioinspired}
\bibinfo{author}{\bibfnamefont{T.-S.} \bibnamefont{Wong}},
  \bibinfo{author}{\bibfnamefont{S.~H.} \bibnamefont{Kang}},
  \bibinfo{author}{\bibfnamefont{S.~K.} \bibnamefont{Tang}},
  \bibinfo{author}{\bibfnamefont{E.~J.} \bibnamefont{Smythe}},
  \bibinfo{author}{\bibfnamefont{B.~D.} \bibnamefont{Hatton}},
  \bibinfo{author}{\bibfnamefont{A.}~\bibnamefont{Grinthal}}, \bibnamefont{and}
  \bibinfo{author}{\bibfnamefont{J.}~\bibnamefont{Aizenberg}},
  \bibinfo{journal}{Nature} \textbf{\bibinfo{volume}{477}},
  \bibinfo{pages}{443} (\bibinfo{year}{2011}{\natexlab{b}}).

\bibitem[{\citenamefont{Chen et~al.}(1999)\citenamefont{Chen, Fadeev, Hsieh,
  {\"O}ner, Youngblood, and McCarthy}}]{chen1999ultrahydrophobic}
\bibinfo{author}{\bibfnamefont{W.}~\bibnamefont{Chen}},
  \bibinfo{author}{\bibfnamefont{A.~Y.} \bibnamefont{Fadeev}},
  \bibinfo{author}{\bibfnamefont{M.~C.} \bibnamefont{Hsieh}},
  \bibinfo{author}{\bibfnamefont{D.}~\bibnamefont{{\"O}ner}},
  \bibinfo{author}{\bibfnamefont{J.}~\bibnamefont{Youngblood}},
  \bibnamefont{and} \bibinfo{author}{\bibfnamefont{T.~J.}
  \bibnamefont{McCarthy}}, \bibinfo{journal}{Langmuir}
  \textbf{\bibinfo{volume}{15}}, \bibinfo{pages}{3395} (\bibinfo{year}{1999}).

\bibitem[{\citenamefont{Tang et~al.}(2020)\citenamefont{Tang, Yao, Tadmor, and
  Sebastian}}]{tang2020lateral}
\bibinfo{author}{\bibfnamefont{S.}~\bibnamefont{Tang}},
  \bibinfo{author}{\bibfnamefont{C.-W.} \bibnamefont{Yao}},
  \bibinfo{author}{\bibfnamefont{R.}~\bibnamefont{Tadmor}}, \bibnamefont{and}
  \bibinfo{author}{\bibfnamefont{D.}~\bibnamefont{Sebastian}},
  \bibinfo{journal}{MRS Communications} \textbf{\bibinfo{volume}{10}},
  \bibinfo{pages}{449} (\bibinfo{year}{2020}).

\bibitem[{\citenamefont{Qu{\'e}r{\'e}}(2005)}]{quere2005non}
\bibinfo{author}{\bibfnamefont{D.}~\bibnamefont{Qu{\'e}r{\'e}}},
  \bibinfo{journal}{Reports on Progress in Physics}
  \textbf{\bibinfo{volume}{68}}, \bibinfo{pages}{2495} (\bibinfo{year}{2005}).

\bibitem[{\citenamefont{Backholm et~al.}(2020)\citenamefont{Backholm,
  Molpeceres, Vuckovac, Nurmi, Hokkanen, Jokinen, Timonen, and
  Ras}}]{backholm2020water}
\bibinfo{author}{\bibfnamefont{M.}~\bibnamefont{Backholm}},
  \bibinfo{author}{\bibfnamefont{D.}~\bibnamefont{Molpeceres}},
  \bibinfo{author}{\bibfnamefont{M.}~\bibnamefont{Vuckovac}},
  \bibinfo{author}{\bibfnamefont{H.}~\bibnamefont{Nurmi}},
  \bibinfo{author}{\bibfnamefont{M.~J.} \bibnamefont{Hokkanen}},
  \bibinfo{author}{\bibfnamefont{V.}~\bibnamefont{Jokinen}},
  \bibinfo{author}{\bibfnamefont{J.~V.} \bibnamefont{Timonen}},
  \bibnamefont{and} \bibinfo{author}{\bibfnamefont{R.~H.} \bibnamefont{Ras}},
  \bibinfo{journal}{Communications Materials} \textbf{\bibinfo{volume}{1}},
  \bibinfo{pages}{1} (\bibinfo{year}{2020}).

\bibitem[{\citenamefont{Extrand and Gent}(1990)}]{extrand1990retention}
\bibinfo{author}{\bibfnamefont{C.}~\bibnamefont{Extrand}} \bibnamefont{and}
  \bibinfo{author}{\bibfnamefont{A.}~\bibnamefont{Gent}},
  \bibinfo{journal}{Journal of Colloid and Interface Science}
  \textbf{\bibinfo{volume}{138}}, \bibinfo{pages}{431} (\bibinfo{year}{1990}).

\bibitem[{\citenamefont{Daniel et~al.}(2018)\citenamefont{Daniel, Timonen, Li,
  Velling, Kreder, Tetreault, and Aizenberg}}]{DanielPRL2018}
\bibinfo{author}{\bibfnamefont{D.}~\bibnamefont{Daniel}},
  \bibinfo{author}{\bibfnamefont{J.~V.~I.} \bibnamefont{Timonen}},
  \bibinfo{author}{\bibfnamefont{R.}~\bibnamefont{Li}},
  \bibinfo{author}{\bibfnamefont{S.~J.} \bibnamefont{Velling}},
  \bibinfo{author}{\bibfnamefont{M.~J.} \bibnamefont{Kreder}},
  \bibinfo{author}{\bibfnamefont{A.}~\bibnamefont{Tetreault}},
  \bibnamefont{and}
  \bibinfo{author}{\bibfnamefont{J.}~\bibnamefont{Aizenberg}},
  \bibinfo{journal}{Phys. Rev. Lett.} \textbf{\bibinfo{volume}{120}},
  \bibinfo{pages}{244503} (\bibinfo{year}{2018}),
  \urlprefix\url{https://link.aps.org/doi/10.1103/PhysRevLett.120.244503}.

\bibitem[{\citenamefont{Liu et~al.}(2019)\citenamefont{Liu, Vuckovac, Latikka,
  Huhtamäki, and Ras}}]{Liu2019Science}
\bibinfo{author}{\bibfnamefont{K.}~\bibnamefont{Liu}},
  \bibinfo{author}{\bibfnamefont{M.}~\bibnamefont{Vuckovac}},
  \bibinfo{author}{\bibfnamefont{M.}~\bibnamefont{Latikka}},
  \bibinfo{author}{\bibfnamefont{T.}~\bibnamefont{Huhtamäki}},
  \bibnamefont{and} \bibinfo{author}{\bibfnamefont{R.~H.~A.}
  \bibnamefont{Ras}}, \bibinfo{journal}{Science}
  \textbf{\bibinfo{volume}{363}}, \bibinfo{pages}{1147} (\bibinfo{year}{2019}),
  \urlprefix\url{https://www.science.org/doi/abs/10.1126/science.aav5388}.

\bibitem[{\citenamefont{Furmidge}(1962)}]{furmidge1962studies}
\bibinfo{author}{\bibfnamefont{C.}~\bibnamefont{Furmidge}},
  \bibinfo{journal}{Journal of colloid science} \textbf{\bibinfo{volume}{17}},
  \bibinfo{pages}{309} (\bibinfo{year}{1962}).

\bibitem[{\citenamefont{ElSherbini and Jacobi}(2006)}]{elsherbini2006retention}
\bibinfo{author}{\bibfnamefont{A.}~\bibnamefont{ElSherbini}} \bibnamefont{and}
  \bibinfo{author}{\bibfnamefont{A.}~\bibnamefont{Jacobi}},
  \bibinfo{journal}{Journal of colloid and interface science}
  \textbf{\bibinfo{volume}{299}}, \bibinfo{pages}{841} (\bibinfo{year}{2006}).

\bibitem[{\citenamefont{De~Gennes et~al.}(2004)\citenamefont{De~Gennes,
  Brochard-Wyart, Qu{\'e}r{\'e} et~al.}}]{de2004capillarity}
\bibinfo{author}{\bibfnamefont{P.-G.} \bibnamefont{De~Gennes}},
  \bibinfo{author}{\bibfnamefont{F.}~\bibnamefont{Brochard-Wyart}},
  \bibinfo{author}{\bibfnamefont{D.}~\bibnamefont{Qu{\'e}r{\'e}}},
  \bibnamefont{et~al.}, \emph{\bibinfo{title}{Capillarity and wetting
  phenomena: drops, bubbles, pearls, waves}}, vol. \bibinfo{volume}{315}
  (\bibinfo{publisher}{Springer}, \bibinfo{year}{2004}).

\bibitem[{\citenamefont{Holly and Refojo}(1975)}]{Holly1975WettabilityOH}
\bibinfo{author}{\bibfnamefont{F.~J.} \bibnamefont{Holly}} \bibnamefont{and}
  \bibinfo{author}{\bibfnamefont{M.~F.} \bibnamefont{Refojo}},
  \bibinfo{journal}{Journal of biomedical materials research}
  \textbf{\bibinfo{volume}{9 3}}, \bibinfo{pages}{315} (\bibinfo{year}{1975}).

\bibitem[{\citenamefont{Schroen et~al.}(1995)\citenamefont{Schroen, Stuart,
  Van~der Voort~Maarschalk, Van~der Padt, and
  Van't~Riet}}]{schroen1995influence}
\bibinfo{author}{\bibfnamefont{C.}~\bibnamefont{Schroen}},
  \bibinfo{author}{\bibfnamefont{M.~C.} \bibnamefont{Stuart}},
  \bibinfo{author}{\bibfnamefont{K.}~\bibnamefont{Van~der Voort~Maarschalk}},
  \bibinfo{author}{\bibfnamefont{A.}~\bibnamefont{Van~der Padt}},
  \bibnamefont{and}
  \bibinfo{author}{\bibfnamefont{K.}~\bibnamefont{Van't~Riet}},
  \bibinfo{journal}{Langmuir} \textbf{\bibinfo{volume}{11}},
  \bibinfo{pages}{3068} (\bibinfo{year}{1995}).

\bibitem[{\citenamefont{Schmidt and Scherzer}(2015)}]{schmidt2015monitoring}
\bibinfo{author}{\bibfnamefont{C.}~\bibnamefont{Schmidt}} \bibnamefont{and}
  \bibinfo{author}{\bibfnamefont{T.}~\bibnamefont{Scherzer}},
  \bibinfo{journal}{Journal of Polymer Science Part B: Polymer Physics}
  \textbf{\bibinfo{volume}{53}}, \bibinfo{pages}{729} (\bibinfo{year}{2015}).

\bibitem[{\citenamefont{Law and Zhao}(2016)}]{Law2016SurfaceWC}
\bibinfo{author}{\bibfnamefont{K.-Y.} \bibnamefont{Law}} \bibnamefont{and}
  \bibinfo{author}{\bibfnamefont{H.}~\bibnamefont{Zhao}},
  \bibinfo{journal}{Surface Wetting}  (\bibinfo{year}{2016}).

\bibitem[{\citenamefont{Cassie and Baxter}(1944)}]{Cassie1944WettabilityOP}
\bibinfo{author}{\bibfnamefont{A.~B.~D.} \bibnamefont{Cassie}}
  \bibnamefont{and} \bibinfo{author}{\bibfnamefont{S.}~\bibnamefont{Baxter}},
  \bibinfo{journal}{Transactions of The Faraday Society}
  \textbf{\bibinfo{volume}{40}}, \bibinfo{pages}{546} (\bibinfo{year}{1944}).

\bibitem[{\citenamefont{DeLong et~al.}(2005)\citenamefont{DeLong, Gobin, and
  West}}]{delong2005covalent}
\bibinfo{author}{\bibfnamefont{S.~A.} \bibnamefont{DeLong}},
  \bibinfo{author}{\bibfnamefont{A.~S.} \bibnamefont{Gobin}}, \bibnamefont{and}
  \bibinfo{author}{\bibfnamefont{J.~L.} \bibnamefont{West}},
  \bibinfo{journal}{Journal of Controlled Release}
  \textbf{\bibinfo{volume}{109}}, \bibinfo{pages}{139} (\bibinfo{year}{2005}).

\bibitem[{\citenamefont{Yuk et~al.}(2016)\citenamefont{Yuk, Zhang, Lin, Parada,
  and Zhao}}]{yuk2016tough}
\bibinfo{author}{\bibfnamefont{H.}~\bibnamefont{Yuk}},
  \bibinfo{author}{\bibfnamefont{T.}~\bibnamefont{Zhang}},
  \bibinfo{author}{\bibfnamefont{S.}~\bibnamefont{Lin}},
  \bibinfo{author}{\bibfnamefont{G.~A.} \bibnamefont{Parada}},
  \bibnamefont{and} \bibinfo{author}{\bibfnamefont{X.}~\bibnamefont{Zhao}},
  \bibinfo{journal}{Nature materials} \textbf{\bibinfo{volume}{15}},
  \bibinfo{pages}{190} (\bibinfo{year}{2016}).

\bibitem[{\citenamefont{Jeon et~al.}(2020)\citenamefont{Jeon, Nguyen, and
  Saleh}}]{Jeon2020JPCB}
\bibinfo{author}{\bibfnamefont{B.-j.} \bibnamefont{Jeon}},
  \bibinfo{author}{\bibfnamefont{D.~T.} \bibnamefont{Nguyen}},
  \bibnamefont{and} \bibinfo{author}{\bibfnamefont{O.~A.} \bibnamefont{Saleh}},
  \bibinfo{journal}{The Journal of Physical Chemistry B}
  \textbf{\bibinfo{volume}{124}}, \bibinfo{pages}{8888} (\bibinfo{year}{2020}),
  \bibinfo{note}{pMID: 32960601},
  \urlprefix\url{https://doi.org/10.1021/acs.jpcb.0c06911}.

\bibitem[{\citenamefont{Angelova and Dimitrov}(1986)}]{Angelova}
\bibinfo{author}{\bibfnamefont{M.~I.} \bibnamefont{Angelova}} \bibnamefont{and}
  \bibinfo{author}{\bibfnamefont{D.~S.} \bibnamefont{Dimitrov}},
  \bibinfo{journal}{Faraday Discuss. Chem. Soc.} \textbf{\bibinfo{volume}{81}},
  \bibinfo{pages}{303} (\bibinfo{year}{1986}),
  \urlprefix\url{http://dx.doi.org/10.1039/DC9868100303}.

\bibitem[{\citenamefont{Angelova et~al.}(1992)\citenamefont{Angelova,
  Sol{\'e}au, M{\'e}l{\'e}ard, Faucon, and Bothorel}}]{Angelova02}
\bibinfo{author}{\bibfnamefont{M.~I.} \bibnamefont{Angelova}},
  \bibinfo{author}{\bibfnamefont{S.}~\bibnamefont{Sol{\'e}au}},
  \bibinfo{author}{\bibfnamefont{P.}~\bibnamefont{M{\'e}l{\'e}ard}},
  \bibinfo{author}{\bibfnamefont{F.}~\bibnamefont{Faucon}}, \bibnamefont{and}
  \bibinfo{author}{\bibfnamefont{P.}~\bibnamefont{Bothorel}}, in
  \emph{\bibinfo{booktitle}{Trends in Colloid and Interface Science VI}},
  edited by \bibinfo{editor}{\bibfnamefont{C.}~\bibnamefont{Helm}},
  \bibinfo{editor}{\bibfnamefont{M.}~\bibnamefont{L{\"o}sche}},
  \bibnamefont{and}
  \bibinfo{editor}{\bibfnamefont{H.}~\bibnamefont{M{\"o}hwald}}
  (\bibinfo{publisher}{Steinkopff}, \bibinfo{address}{Darmstadt},
  \bibinfo{year}{1992}), pp. \bibinfo{pages}{127--131}.

\bibitem[{\citenamefont{Schneider et~al.}(2012)\citenamefont{Schneider,
  Rasband, and Eliceiri}}]{ImageJ}
\bibinfo{author}{\bibfnamefont{C.~A.} \bibnamefont{Schneider}},
  \bibinfo{author}{\bibfnamefont{W.~S.} \bibnamefont{Rasband}},
  \bibnamefont{and} \bibinfo{author}{\bibfnamefont{K.~W.}
  \bibnamefont{Eliceiri}}, \bibinfo{journal}{Nature Methods}
  \textbf{\bibinfo{volume}{9}}, \bibinfo{pages}{671} (\bibinfo{year}{2012}),
  \urlprefix\url{https://doi.org/10.1038/nmeth.2089}.

\bibitem[{\citenamefont{Hell et~al.}(1993)\citenamefont{Hell, Reiner, Cremer,
  and Stelzer}}]{hell1993aberrations}
\bibinfo{author}{\bibfnamefont{S.}~\bibnamefont{Hell}},
  \bibinfo{author}{\bibfnamefont{G.}~\bibnamefont{Reiner}},
  \bibinfo{author}{\bibfnamefont{C.}~\bibnamefont{Cremer}}, \bibnamefont{and}
  \bibinfo{author}{\bibfnamefont{E.~H.} \bibnamefont{Stelzer}},
  \bibinfo{journal}{Journal of microscopy} \textbf{\bibinfo{volume}{169}},
  \bibinfo{pages}{391} (\bibinfo{year}{1993}).

\bibitem[{\citenamefont{Sheppard and Torok}(1997)}]{sheppard1997effects}
\bibinfo{author}{\bibfnamefont{C.}~\bibnamefont{Sheppard}} \bibnamefont{and}
  \bibinfo{author}{\bibfnamefont{P.}~\bibnamefont{Torok}},
  \bibinfo{journal}{Journal of microscopy} \textbf{\bibinfo{volume}{185}},
  \bibinfo{pages}{366} (\bibinfo{year}{1997}).

\bibitem[{\citenamefont{R{\'i}o and Neumann}(1997)}]{Ro1997AxisymmetricDS}
\bibinfo{author}{\bibnamefont{R{\'i}o}} \bibnamefont{and}
  \bibinfo{author}{\bibnamefont{Neumann}}, \bibinfo{journal}{Journal of colloid
  and interface science} \textbf{\bibinfo{volume}{196 2}}, \bibinfo{pages}{136}
  (\bibinfo{year}{1997}).

\end{thebibliography}

\section{Supporting Information}
\label{sec:supporting_information}

\subsection{Dynamic contact angle}
\begin{figure}
   \vspace{0.05cm}
   \includegraphics[width=1\columnwidth]{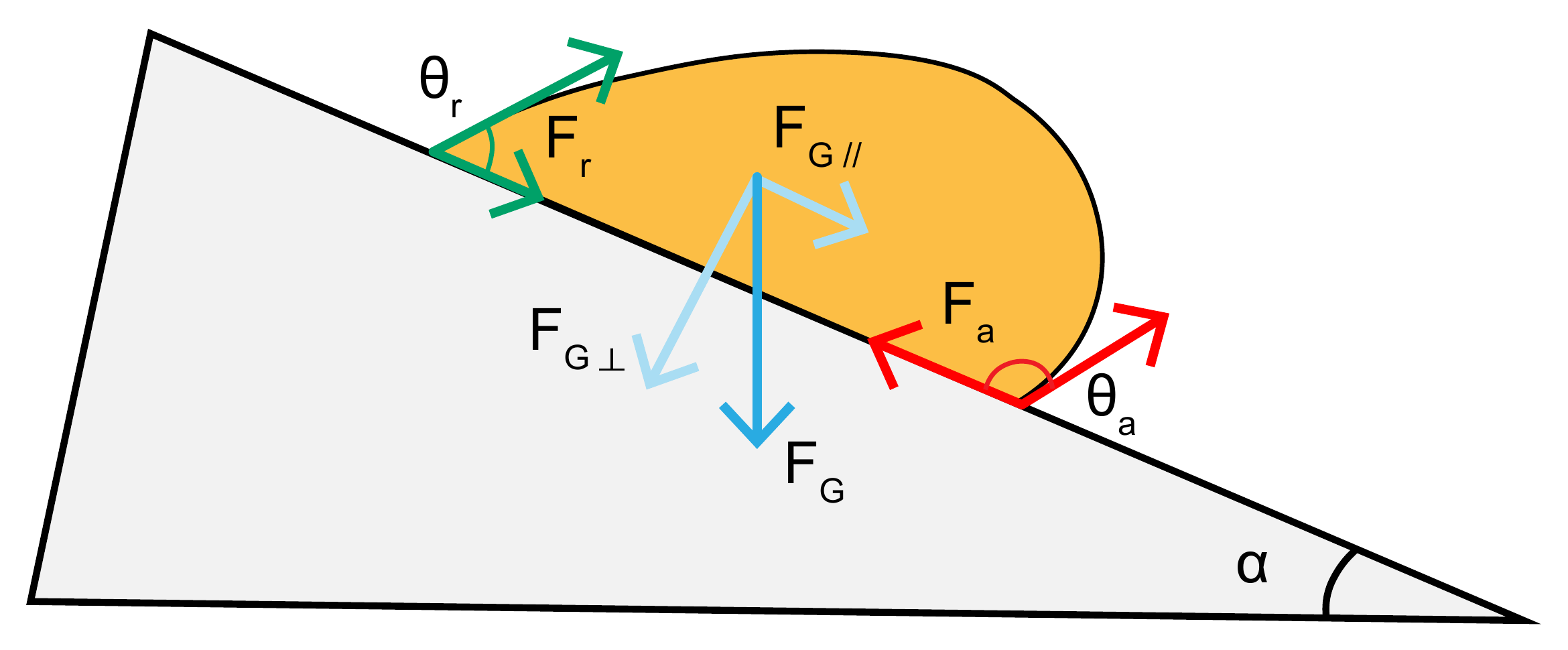}
    \caption{\textit{Supporting Figure: Force balance of a droplet sliding on a tilted surface.}
    With advancing contact angle $\theta_a$ and receding contact angle $\theta_r$
    }
    \vspace{0.05cm}
    \label{fig:theory_sliding_drop}
\end{figure}

Contact angle hysteresis is the difference between the contact angle of a droplet sitting on a surface in wetting and dewetting regions.
While on perfectly defect-free, ideal surfaces the contact angle $\theta$ is a unique value, on a real surface and in dynamic conditions, $\theta$ varies between two extremes, $\theta_a$ and $\theta_r$, known as the advancing and receding contact angles, respectively \cite{shi2018dynamic, eral2013contact}. If we imagine a droplet sitting on a flat surface, its shape will be symmetrical.
Let's now tilt the surface at an angle $\alpha$.
There is a force, the \emph{pinning force}, between the droplet and the surface, which will maintain the droplet adhered to the surface.
Hence, it will initially not move, but it will start deforming under its own weight (Fig. S\ref{fig:theory_sliding_drop}).
When a critical value of $\alpha$ is reached, however, the droplet will start sliding.
Because of the deformation, the contact angles at the advancing and receding ends will be different: these are $\theta_a$ and $\theta_r$.
The stronger the pinning force, the more the droplet will deform before starting to move, hence the larger the difference $\Delta\theta=\theta_a-\theta_r$.

The values of $\theta_a$ and $\theta_r$ can be calculated by a simple force balance \cite{de2004capillarity}.
A side-view of the droplet is represented in Fig. S\ref{fig:theory_sliding_drop}. 
The droplet slides as a result of the parallel component of the gravitational force with respect to the surface, $F_{G\parallel}$, which be expressed as
\begin{equation}\label{eq:grav_force}
    F_{G\parallel} = \Delta \rho \cdot V \cdot g \cdot \sin\alpha,
\end{equation}
where $\Delta \rho$ is the density difference between the droplet and the surrounding medium, $V$ the droplet volume, $g$ the gravitational acceleration, and $\alpha$ the tilting angle.

The pinning force $F$, on the other hand, counteracts droplet motion.
This should be derived in principle from the droplet surface tension balance along the whole droplet contact line.
In most studies, however, its value is simply determined by the balance of forces at the advancing ($f_a$) and receding ($f_r$) ends, multiplied by a geometrical pre-factor, $k$:
\begin{equation}\label{eq:theory_forcebalance1}
    F = k \cdot (f_r - f_a).
\end{equation}
Several values of $k$ have been reported in literature \cite{wong2011bioinspired}.
For droplets with circular contact area, $k=\frac{48}{\pi^3}R$ offers a good approximation,  where $R$ is the droplet radius as seen from the side \cite{elsherbini2006retention}.
By expressing $f_a$ and $f_r$ in terms of the droplet surface tension $\gamma$, we can then rewrite Eq. \ref{eq:theory_forcebalance1} as
\begin{equation}\label{eq:retention_force}
    F = \frac{48R}{\pi^3} \cdot ( \gamma \cos\theta_r -  \gamma \cos\theta_a).
\end{equation}
At the onset of motion, $F_{G \parallel} = F$, hence:
\begin{equation}
    \Delta \rho \cdot g \cdot V  \cdot \sin\alpha = \frac{48R}{\pi^3}\cdot\gamma \cdot(\cos\theta_r - \cos\theta_a),
\end{equation}
which then can be rearranged as
\begin{equation}\label{eq:force_balance2}
    (\cos\theta_r - \cos\theta_a) = \frac{\Delta\rho \cdot g \cdot V\cdot{\pi^3}}{48R\cdot\gamma}\cdot \sin\alpha.
\end{equation}
Importantly, we can estimate the retention force using Eq. \ref{eq:grav_force} by measuring the critical tilting angle $\alpha$, the droplet volume $V$, the density difference $\Delta\rho$, and knowing the gravitational acceleration $g$. Equation \ref{eq:force_balance2} can be simplified using Taylor expansion of $\cos\theta$ around $\theta\approx180^\circ$ to write
\begin{equation}\label{eq:taylor}
    (\cos\theta_r - \cos\theta_a)\approx\pi\cdot\Delta\theta,
\end{equation}
and also using $V\approx\frac{4\pi R^3}{3}$, to yield
\begin{equation}\label{eq:force_balance3}
    \Delta\theta\approx\frac{\Delta\rho \cdot g \cdot R^2 \cdot \pi^3}{36~\gamma}\cdot \sin\alpha.
\end{equation}
Here, the relevant dimensionless quantity is the Bond (or Eötvös) number:
\begin{equation}
    Bo=\frac{\Delta\rho\cdot g\cdot R^2 }{\gamma}.
\end{equation}
Thus,
\begin{equation}\label{eq:hysteresis_final}
    \Delta\theta \approx \frac{\pi^3}{36}\cdot Bo \cdot \sin\alpha,
\end{equation}
that is, the critical tilt angles can only be used as a relative measure of contact hysteresis between droplets with similar Bond numbers.
\subsection{Supporting Figures}
\begin{figure*}
    \centering
   \includegraphics[width=0.9\textwidth]{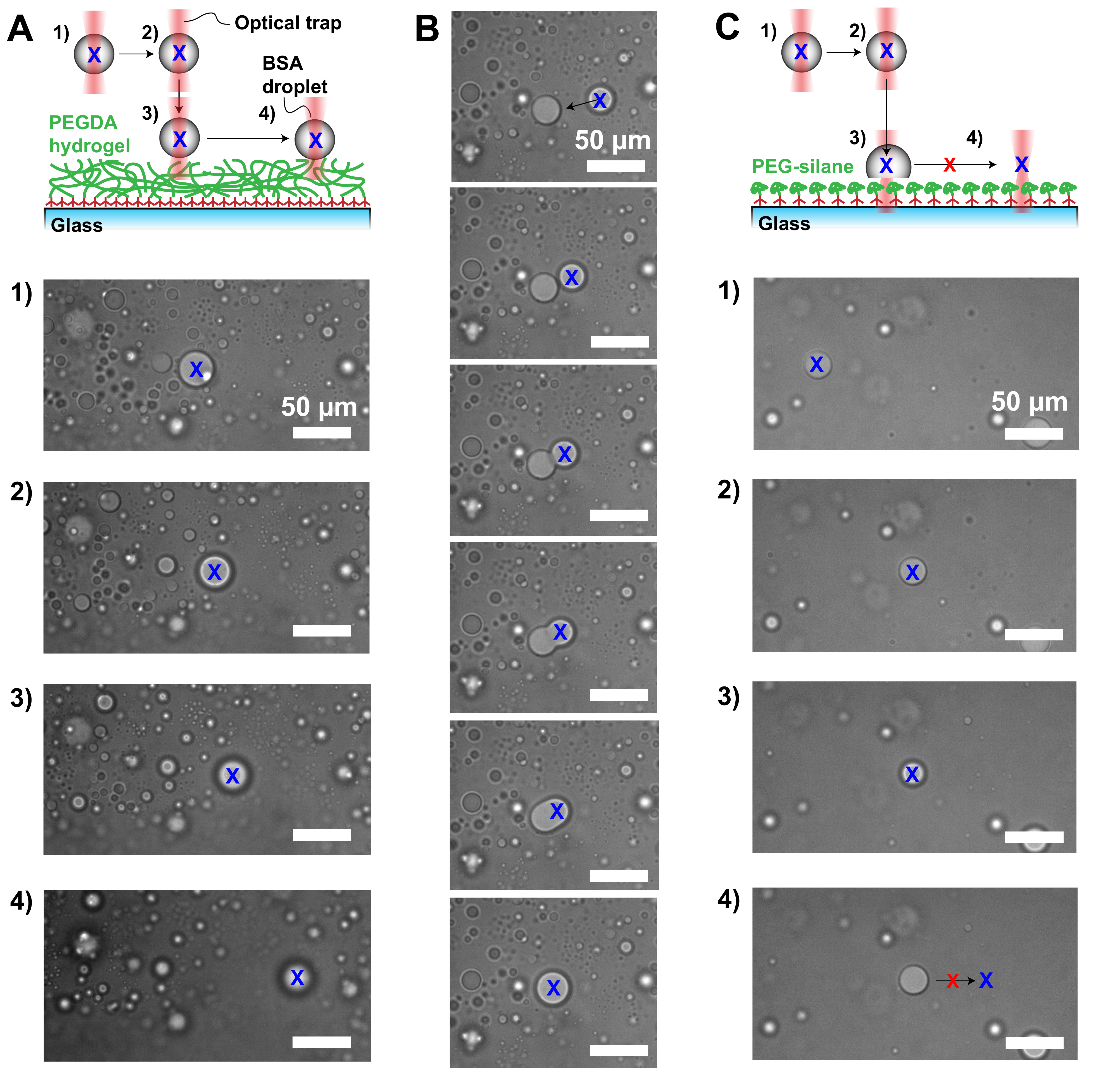}
    \caption{\textit{Supporting Figure: Image sequences for BSA droplet micromanipulation experiments with optical tweezers.} 
    \textbf{(A)} Top, schematic of micromanipulation of BSA droplets (\textit{gray})with an optical trap (\textit{red}) on PEGDA hydrogel (\textit{green}). \textit{1)} we first grabbed a floating droplet from solution, \textit{2)} we then moved it to make sure the droplet was trapped, \textit{3)} the droplet was then lowered and gently pushed against the hydrogel, \textit{4)}, lastly, we moved the droplet on the hydrogel to verity the absence of pinning. Bottom, bright field images corresponding to experimental sequence \textit{1)}-\textit{4)}. The location of the center of the trap is marked as a blue 'X'.
    \textbf{(B)} Bright field image sequence of a trapped BSA droplet moved to another droplet and their subsequent fusion process.  
    \textbf{(C)} Top, schematic of micromanipulation of BSA droplets (\textit{gray}) with an optical trap (\textit{red}) on PEG-silane (\textit{green}). After trapping \textit{(1)}, moving \textit{(2)}, and lowering \textit{(3)} the droplet, it partially wetted the PEG-silane, and we could no longer move the droplet \textit{(4)}, \textit{(crossed out arrow)}.
    }
    \label{supp_fig:tweezers}
\end{figure*}
\begin{table*}
    \centering
    \caption{AC treatment program for electroformation of GUVs}
    \vspace{0.2cm}
    \begin{tabular}{|c|c|c|c|}
         \hline
         Step & Frequency & Voltage (peak-to-peak) & Time \\
         \hline
        1 & 10 Hz & 830 mV & 5 min \\
        2 & 10 Hz & 1.66 V & 5 min \\
        3 & 10 Hz & 2.5 V & 5 min \\
        4 & 10 Hz & 3.3 V & 5 min \\
        5 & 10 Hz & 4.1 V & 5 min \\
        6 & 10 Hz & 5 V & 2 hours \\
        7 & 5 Hz & 5 V & 30 min \\
        \hline
    \end{tabular}
    \label{supp_tab:electroformation}
\end{table*}
\clearpage
\end{document}